\documentclass[]{aa} 
\usepackage{natbib} 
\usepackage{graphicx}

\def\rms{r{.}m{.}s{.}~}
\def\resp{resp{.}}
\def\ie{i{.}e{.}~}
\def\eq{i{.}e{.}}
\def\eq{Eq{.}}                 
\def\eqs{Eqs{.}}
\def\fg{Fig{.}}
\def\fgs{Figs{.}}
\def\sct{Sect{.}}
\def\scts{Sects{.}}

\def\mag{$^\mathrm{mag}$}

\def\hmpc{~$h^{-1}$~Mpc}

\def\phiunit{~$h^3$~Mpc$^{-3}$~mag$^{-1}$}

\begin{document}

\title{The ESO-Sculptor Survey: Evolution of late-type galaxies 
at redshifts $0.1-0.5$ \thanks{Based on observations collected at the
European Southern Observatory (ESO), La Silla, Chile.}}

\author{Val\'erie de Lapparent\inst{1} 
   \and St\'ephane Arnouts\inst{2}
   \and Gaspar Galaz\inst{3} 
   \and Sandro Bardelli\inst{4}} 

\offprints{lapparen@iap.fr}

\institute{Institut d'Astrophysique de Paris, CNRS, Univ{.} Pierre et Marie Curie, 98 bis Boulevard Arago, 75014 Paris, France\\
\email{lapparen@iap.fr} 
\and Laboratoire d'Astrophysique de Marseille, BP8, Traverse du Siphon, 13376 Marseille Cedex 12, France\\
\email{stephane.arnouts@oamp.fr} 
\and 
Depto. de Astronom\'\i a et Astrof\'\i sica, Pontificia Universidad Cat\'olica de Chile, casilla 306, Santiago 22, Chile\\ 
\email{ggalaz@astro.puc.cl} 
\and
INAF-Osservatorio Astronomico di Bologna, via Ranzani 1, 40127 Bologna, Italy\\ 
\email{bardelli@excalibur.bo.astro.it} }

\date{Received 27 January 2004 / Accepted 1 April 2004}

\abstract{

Using the Gaussian+Schechter composite luminosity functions measured
from the ESO-Sculptor Survey \citep{lapparent03aI}, and assuming that
these functions do not evolve with redshift out to $z\sim1$, we obtain
evidence for evolution in the late spectral class containing late-type
Spiral (Sc+Sd) and dwarf Irregular (dI) galaxies. There are
indications that the Sc+Sd galaxies are the evolving population, but
we cannot exclude that the dI galaxies also undergo some evolution.
This evolution is detected as an increase of the Sc+Sd+Im galaxy
density which can be modeled as either $n(z)\propto1+3(z-0.15)$ or
$n(z)\propto(1+z)^2$ using the currently favored cosmological
parameters $\Omega_m=0.3$ and $\Omega_\Lambda=0.7$; the uncertainty in
the linear and power-law evolution rates is of order of unity.  For
$\Omega_m=1.0$ and $\Omega_\Lambda=0.0$, the linear and power-law
evolution rates are $\sim4\pm1$ and $\sim2.5\pm1$ respectively.  Both
models yield a good match to the ESS $BVR_\mathrm{c}$ redshift
distributions to $21-22$\mag~and to the number-counts to
$23-23.5$\mag, which probe the galaxy distribution to redshifts
$z\sim0.5$ and $z\sim1.0$ respectively.

\hspace{0.5cm} 
 
The present analysis shows the usefulness of the joint use of the
magnitude and redshift distributions for studying galaxy evolution. It
also illustrates how Gaussian+Schechter composite luminosity functions
provide more robust constraints on the evolution rate than pure
Schechter luminosity functions, thus emphasizing the importance of
performing realistic parameterizations of the luminosity functions for
studying galaxy evolution.

\hspace{0.5cm} 
 
The detected density evolution indicates that mergers could play
a significant role in the evolution of late-type Spiral and dwarf
Irregular galaxies. However, the ESO-Sculptor density increase with
redshift could also be caused by a $\sim1$\mag~brightening of the
Sc+Sd+dI galaxies at $z\sim0.5$ and a $\sim1.5-2.0$\mag~brightening at
$z\sim1$, which is compatible with the expected passive brightening of
Sc galaxies at these redshifts. Distinguishing between luminosity and
density evolution is a major difficulty as these produce the same
effect on the redshift and magnitude distributions.  The detected
evolution rate of the ESO-Sculptor Sc+Sd+dI galaxies is nevertheless among the
range of measured values from the other existing analyses, whether
they provide evidence for density or luminosity evolution.


}

\authorrunning{de Lapparent et al.}
\titlerunning{The ESO-Sculptor Survey, Evolution at $z\simeq0.1-0.5$}
\maketitle

\section{Introduction \label{intro}} 

Since the availability of the deep optical numbers counts, the excess
at faint magnitudes has provided the major evidence for galaxy
evolution at increasing redshifts \citep{tyson88,lilly93,metcalfe95}.
Using models of the spectro-photometric evolution of galaxies
\citep{guiderdoni90,bruzual93}, either passive luminosity evolution or
more complex effects have been suggested to explain the faint
number-count excess
\citep{guiderdoni91,broadhurst92,metcalfe95}. Although the excess
objects where initially envisioned as bright early-type galaxies at
high redshift, the lack of a corresponding high redshift tail in the
redshift distribution \citep{lilly93b} consolidated the interpretation
in terms of evolution of later type galaxies, namely Spiral and/or
Irregular/Peculiar galaxies \citep{campos97b}.

Here, we report on yet another evidence for evolution of the late-type
galaxies, derived from the ESO-Sculptor Survey (ESS hereafter). The
ESS provides a nearly complete redshift survey of galaxies at
$z\la0.5$ over a contiguous area of the sky.  A reliable description
of galaxy evolution require proper identification of the evolving
galaxy populations and detailed knowledge of their luminosity
functions.  In this context, the ESS sample has the advantage to be
split into 3 galaxy classes which are based on a template-free
spectral classification \citep{galaz98}, and which are dominated by
the giant morphological types E+S0+Sa, Sb+Sc, and Sc+Sd+Sm
respectively \citep[][Paper~I hereafter]{lapparent03aI}. In Paper~I,
we have performed a detailed measurement of the shape of the
luminosity functions (LF hereafter) for the 3 ESS spectral classes.
The spectral-type LFs show marked differences among the classes, which
are common to the $B$, $V$, $R_\mathrm{c}$ bands, and thus indicate
that they measure physical properties of the underlying galaxy
populations.

The analysis of the ESS LFs in Paper~I also provides a revival of the
view advocated by \citet{binggeli88}: a galaxy LF is the weighted
sum of the \emph{intrinsic} LFs for each morphological type contained
in the considered galaxy sample; in this picture, differences in LFs
mark variations in the galaxy mix rather than variations in the
intrinsic LFs
\citep{dressler80,postman84,binggeli90,ferguson91,trentham02a,trentham02b}.
Local measures show that giant galaxies (Elliptical, Lenticular, and
Spiral) have Gaussian LFs, which are thus bounded at both bright and
faint magnitudes, with the Elliptical LF skewed towards faint
magnitudes \citep{sandage85b,jerjen97b}. In contrast, the LF for dwarf
Spheroidal galaxies may be ever increasing at faint magnitudes to the
limit of the existing surveys
\citep{sandage85b,ferguson91,jerjen00,flint01a,flint01b,conselice02},
whereas the LF for dwarf Irregular galaxies is flatter
\citep{pritchet99} and may even be bounded at faint magnitudes
\citep{ferguson89b,jerjen97b,jerjen00}. In Paper~I, by fitting the ESS
spectral-type LFs with composite functions based on the Gaussian and
Schechter LFs measured for each morphological type in local galaxy
groups and clusters \citep{sandage85b,jerjen97b}, we confirm the
morphological content in giant galaxies of the ESS classes, and we
detect an additional contribution from dwarf Spheroidal (dE) and dwarf
Irregular galaxies (dI) in the intermediate-type and late-type classes
respectively.  We then suggest that by providing a good match to the
ESS spectral-type LFs, the local intrinsic LFs may extend to
$z\sim0.5$ with only small variations.

In the following, we report on the measurement of the amplitude of the
LFs for the 3 ESS spectral-type LFs, and on the detection and
measurement of redshift evolution for the late-type galaxies.
\sct~\ref{ess} lists the main characteristics of the ESS spectroscopic
survey.  \sct~\ref{sp_kcor} recalls the definition of the ESS spectral
classes and the technique for deriving the corresponding K-corrections
and absolute magnitudes.  \sct~\ref{shape} shows the measured
composite fits of the ESS LFs for the 3 spectral classes. In
\sct~\ref{phistar}, we describe the various techniques for measuring
the amplitude of the LF (\sct~\ref{phistar_equat}) and the associated
errors (\sct~\ref{errors}); we then apply these techniques to the ESS
and show the detected evolution in the late-type galaxies
(\sct~\ref{evol}).  In \sct~\ref{counts}, we use the ESS magnitude
number-counts to derive improved estimate of the late-type galaxy
evolution rate in the $B$, $V$, and $R_\mathrm{c}$ bands.  We then
examine in \sct~\ref{nz} the redshift distributions for the 3 spectral
classes in the 3 filters, and we verify that the measured evolution
rates for the late-type galaxies match the ESS expected redshift
distributions. Then, in \sct~\ref{evol_comp}, we compare the detected
evolution in the ESS LF with those derived from other existing
redshift surveys which detect either number density evolution
(\sct~\ref{dens_comp}) or luminosity evolution (\sct~\ref{lum_comp}).
Finally, \sct~\ref{concl} summarizes the results, discusses them in
view of the other analyses which detect evolution in the late-type
galaxies, and raises some of the prospects.

\section{The ESS spectroscopic survey \label{ess}}

The ESO-Sculptor Survey (ESS hereafter) provides a complete
photometric and spectroscopic survey of galaxies in a region centered
at $\sim 0^\mathrm{h}22^\mathrm{m}$ (R{.}A{.}) $\sim
-30^\circ06^\prime$ (DEC{.}), near the Southern Galactic Pole.  The
photometric survey provides standard magnitudes $B$, $V$, and
$R_\mathrm{c}$ in the Johnson-Cousins system, for nearly 13000
galaxies to $V \simeq 24$ over a contiguous rectangular area of $\sim
0.37$ deg$^2$ [$1.53^\circ\mathrm{(R{.}A{.})}  \times
0.24^\circ\mathrm{(DEC{.})}$] \citep{arnouts97}.  The uncertainties in
the apparent magnitudes are $0.05$\mag~in the $B$ $V$ and
$R_\mathrm{c}$ bands for $R\mathrm{c}\la 21.0$ \citep{arnouts97}.
Multi-slit spectroscopy of the $\sim 600$ galaxies with $R_\mathrm{c}
\le 20.5$ \citep{bellanger95a} have provided a 92\% complete redshift
survey over a contiguous sub-area of $\sim 0.25$ deg$^2$
[$1.02^\circ\mathrm{(R{.}A{.})}  \times 0.24^\circ\mathrm{(DEC{.})}$].
Additional redshifts for $\sim 250$ galaxies with $20.5< R_\mathrm{c}
\le 21.5$ were also measured in the same sub-area, leading to a 52\%
redshift completeness to $R_\mathrm{c} \le 21.5$ (see Paper~I for
details). We also consider here the $V\le21.0$ and $B\le22.0$
redshifts samples, which correspond to the combination of the
$R_\mathrm{c}\le20.5$ ``nominal'' limit with the typical colors of
galaxies at that limit: $B-R_\mathrm{c}\simeq1.5$ and $V-R_\mathrm{c}
\simeq0.5$ \citep{arnouts97}). The redshift completeness for the
$V\le21.0$ and $B\le22.0$ samples is $91$\% and $86$\% respectively.

\section{Spectral classification and K-corrections  \label{sp_kcor} }  

Estimates of morphological types are not available for the ESS
redshift survey.  In Paper~I, we estimate the ESS intrinsic LFs based
on a spectral classification.  Using a Principal Component Analysis,
we have derived an objective spectral sequence, which is parameterized
continuously using 2 parameters, describing respectively the relative
fractions of old to young stellar populations (parameter denoted here
$T_\mathrm{S}$), and the relative strength of the emission lines
\citep{galaz98}.

The ESS spectral sequence is separated into 3 classes, denoted
``early-type'', ``intermediate-type'', and ``late-type'', which
correspond to $T_\mathrm{S}\le-5^\circ$, $-5^\circ<
T_\mathrm{S}\le3^\circ$, and $T_\mathrm{S}>3^\circ$ respectively.
These values separate the ESS spectroscopic $R_\mathrm{c}$, $V$, and
$B$ samples into sub-samples with as least 100 galaxies (see
Table~\ref{lf_BVR}).  Given the moderate number of objects in the ESS
spectroscopic sample, these 3 classes thus provide a satisfying
compromise between resolution in spectral-type and signal-to-noise in
the corresponding LFs.

Projection of the Kennicutt (1992) spectra onto the ESS spectral
sequence shows that a tight correspondence with the Hubble
morphological sequence of normal galaxies (\citealp{galaz98};
Paper~I); the ESS early-type class contains predominantly E, S0 and Sa
galaxies, the intermediate-type class, Sb and Sc galaxies, and the
late-type class, Sc, Sd and Sm/Im galaxies. We show that these
spectral classes allow us to detect the respective contributions to
the LF from the Elliptical, Lenticular and Spiral galaxies, and from
the dwarf Spheroidal and Irregular galaxies.

In Paper~I, we estimate the K-correction for the ESS by projecting
templates extracted from the PEGASE\footnote{Projet d'Etude des
GAlaxies par Synth\`ese Evolutive \citep{fioc97}.} spectrophotometric
model of galaxy evolution onto the ESS spectral sequence. A fine mesh
of model spectra with varying redshift and spectral-type are
generated, and K-corrections in the $B$, $V$, and $R_\mathrm{c}$
filters are calculated for each of them; because the model spectra
extend from 2000 \AA~to 10000 \AA, we can derive K-corrections in the
3 bands up to $z\simeq 0.6$, the effective depth of the ESS redshift
survey.  A 2-D polynomial fit to the resulting surface in each filter
provides analytical formul\ae~for the K-corrections as a function of
redshift and spectral type, which are used for the ESS galaxies; these
are plotted in \fg~\ref{kcor_z2} below, where they are compared with
the K-corrections by \citet[][see Paper~I for further details and
comparisons]{coleman80}. For each galaxy, the absolute magnitude $M$
is then derived from the apparent magnitude $m$, the spectral type
$T_\mathrm{S}$ and the redshift $z$ using
\begin{equation}
M(m,z,T_\mathrm{S}) = m - 5\log d_\mathrm{L}(z) - K(z,T_\mathrm{S}) - 25,
\label{abs_mag}
\end{equation}
where $d_\mathrm{L}(z)$ is the luminosity distance in Mpc
\citep{weinberg76}:
\begin{equation}
d_\mathrm{L}(z)= (1+z)\;\frac{c}{H_0}\;\int_0^z[\;\Omega_\mathrm{m}(1+\upsilon)^3+\Omega_\Lambda\ ]^{-\frac{1}{2}}\;\mathrm{d}\upsilon\,
\label{dl}
\end{equation}
Throughout the present analysis, absolute magnitudes are calculated
with a Hubble constant at present epoch written as $H_0 = 100 h$ km
s$^{-1}$ Mpc$^{-1}$.  We also assume a flat Universe with values of
the dimensionless matter density and cosmological constant assigned to
$\Omega_m=0.3$, $\Omega_\Lambda=0.7$ \resp, as currently favored
\citep{riess98,perlmutter99,phillips01,tonry03}; when specified,
$\Omega_m=1.0$ and $\Omega_\Lambda=0.0$ are also considered (see
\scts~\ref{shape} and \ref{counts}).

In Paper~I, we report on all sources of random and systematic errors
which affect the spectral classification, the K-corrections, and the
absolute magnitudes.  By comparison of the 228 pairs of independent
spectra, we measure ``external'' errors in the K-corrections from 0.07
to 0.21 mag, and resulting uncertainties in the absolute magnitudes
from 0.09 to 0.24\mag~(with larger errors in bluer bands for both the
K-corrections and the absolute magnitudes).  From the 228 pairs of
spectra, we also measure an ``external'' \rms uncertainty in the
redshifts of $\sigma\sim0.00055$, which causes negligeable uncertainty
in the absolute magnitudes compared to the other sources of error.

\section{The shape of the ESS luminosity functions \label{shape}}

\begin{table*}
\caption{Parameters of the Gaussian and Schechter components of the
  composite luminosity functions fitted to the ESO-Sculptor
  spectral-type luminosity functions, in the $R_\mathrm{c}$, $V$, and
  $B$ filters.}
\label{lf_BVR}
\begin{center}
\begin{tabular}{lcccccccc}
\hline
\hline
Sample           & Numb{.}  & Morphol. & \multicolumn{3}{c}{Gaussian component} & \multicolumn{2}{c}{Schechter component}   & $\phi_0$ \\
                 & of gal{.} & content  & $M_0-5\log h$   & $\Sigma_a$ & $\Sigma_b$ & $M^*-5\log h$   & $\alpha$ & $\overline{0.4 \ln10\; \phi^*}$ \\
\hline
\multicolumn{9}{c}{\bf early-type galaxies}  \\
\hline
$R_\mathrm{c}\le21.5$ & 291 & E+S0+Sa    & $-20.87\pm0.23$ & $0.84\pm0.24$ & $1.37\pm0.36$ \\
$V\le21.0$            & 156 & E+S0+Sa    & $-20.37$        & $0.84$        & $1.37$        \\
$B\le22.0$            & 108 & E+S0+Sa    & $-19.47$        & $0.84$        & $1.37$        \\
\hline
\multicolumn{9}{c}{\bf intermediate-type galaxies} \\
\hline
$R_\mathrm{c}\le21.5$ & 270 & Sb+Sc/dSph & $-20.27\pm0.21$ & $0.91\pm0.18$ & & $-19.28\pm0.37$ & $-1.53\pm0.33$ & $0.83$ \\
$V\le21.0$            & 169 & Sb+Sc/dSph & $-19.87$        & $0.91$        & & $-18.88$        & $-1.53$        & $0.83$ \\
$B\le22.0$            & 154 & Sb+Sc/dSph & $-19.27$        & $0.91$        & & $-18.28$        & $-1.53$        & $0.83$ \\
\hline
\multicolumn{9}{c}{\bf late-type galaxies} \\
\hline
$R_\mathrm{c}\le21.5$ & 309 & Sc+Sd/dI   & $-19.16\pm0.29$ & $0.97\pm0.13$ & & $-18.12\pm0.22$ & $-0.30$        & $0.1$ \\
$V\le21.0$            & 168 & Sc+Sd/dI   & $-18.76$        & $0.97$        & & $-17.72$        & $-0.30$        & $0.1$ \\
$B\le22.0$            & 190 & Sc+Sd/dI   & $-18.36$        & $0.97$        & & $-17.32$        & $-0.30$        & $0.1$ \\
\hline

\end{tabular}
\smallskip
\\
\end{center}
\begin{list}{}{}
\item[\underline{Notes:}]
\item[-] For the 2-component luminosity functions fitted to the
intermediate-type and late-type galaxies, the morphological content of
the 2 components appear separated by a ``/''.
\item[-] The listed luminosity function parameters correspond to $H_0 =
100 h$ km s$^{-1}$ Mpc$^{-1}$, $\Omega_m=0.3$, and $\Omega_\Lambda=0.7$.
\end{list}
\end{table*}

\begin{figure*}
\centering
\includegraphics[width=17cm]{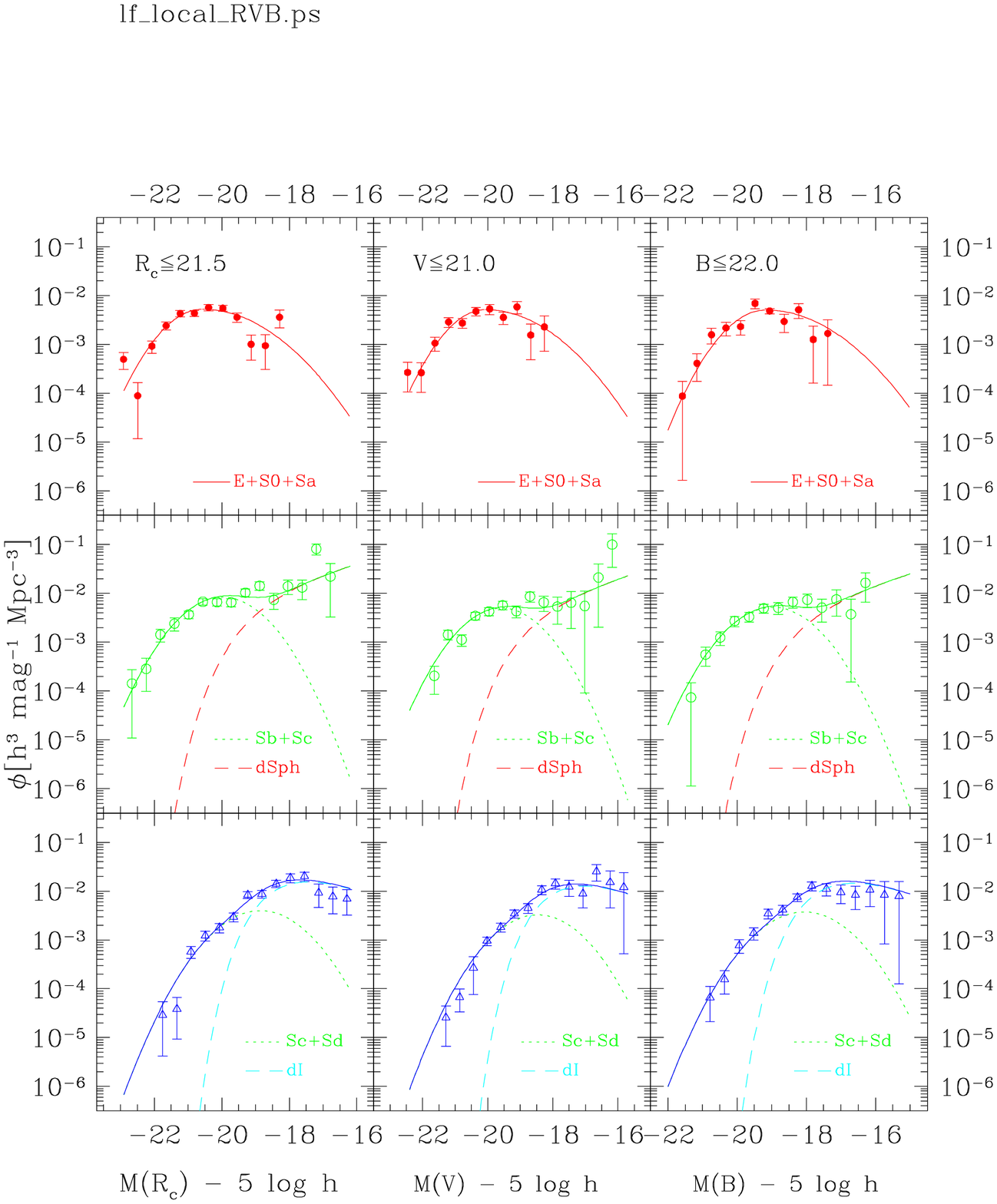}
\caption{The ESO-Sculptor Gaussian+Schechter composite luminosity
functions for the early-type (top panels), intermediate-type (middle
panels), and late-type galaxies (bottom panels) in the 3 samples:
$R_\mathrm{c}\le21.5$ (left panels), $V\le21.0$ (central panels) and
$B\le22.0$ (right panels).  Symbols indicate the SWML solution, and
lines the composite fits obtained; all were obtained with $H_0 = 100
h$ km s$^{-1}$ Mpc$^{-1}$, $\Omega_m=1.0$, and $\Omega_\Lambda=0.0$.
The shape parameters of the composite functions converted to
$\Omega_m=0.3$ and $\Omega_\Lambda=0.7$ are listed in
Table~\ref{lf_BVR}.}
\label{lf_local_RVB}
\end{figure*} 

The shape of the LFs for the 3 ESS spectral classes are derived in
Paper~I, using both the non-parametric step-wise maximum likelihood
method (SWML) developed by \citet{efstathiou88a}, and the method of
\citet[][denote STY]{sandage79} which assumes a specific parametric
form for the LF. Although pure Schechter (1976) functions provide
acceptable STY fits to the 3 ESS spectral-classes, we show that as
good or better STY fits of the $R_\mathrm{c}$ LFs are obtained using
composite functions based on the intrinsic LFs per morphological type
measured in local groups and clusters \citep{jerjen97b,sandage85b}.

For the ESS intermediate-type and late-type $R_\mathrm{c}$ LFs, we fit
the sum of a Gaussian component representing the giant galaxies
(Sb+Sc, and Sc+Sd \resp), and a Schechter component representing the
dwarf galaxies (dSph and dI \resp).  The Gaussian component is
parameterized as
\begin{equation}
\label{gaussian}
\phi(M) \;\mathrm{d}M = \phi_0 e^{-( M_0 - M ) ^2 / 2 \Sigma^2}\;
\mathrm{d}M,
\end{equation}
where $M_0$ and $\Sigma$ are the peak and \rms dispersion respectively.
The \citet{schechter76} component is parameterized as
\begin{equation}
\label{schechter_lum}
\phi(L) \;\mathrm{d}L = \phi^* \left(\frac{L}{L^*}\right)^\alpha
e^{-\frac{L}{L^*}} \;\mathrm{d}\left(\frac{L}{L^*}\right)
\end{equation}
where $\phi^*$ is the the amplitude, $L^*$ the characteristic
luminosity, and $\alpha$ determines the behavior at faint
luminosities.
Rewritten in terms of absolute magnitude, \eq\ref{schechter_lum} becomes:
\begin{equation}
\begin{array}{ll}
\label{schechter_mag}
\phi(M) \;\mathrm{d}M & = 0.4 \ln 10\; \phi^* e^{-X} X^{\alpha+1}\; \mathrm{d}M \\
{\rm with}& \\
X & \equiv \frac{L}{L^*} = 10^{\;0.4\,(M^* - M)} \\
\end{array}
\end{equation}
where $M^*$ is the characteristic magnitude, and $\alpha+1$ the
``faint-end slope''.

For the ESS early-type $R_\mathrm{c}$ LF, a two-wing Gaussian function
is used (a Gaussian with two different dispersion wings at the bright
and faint end), as it successfully reflects the combination of a skewed
LF towards faint magnitudes for the Elliptical galaxies with a
Gaussian LF for the Lenticular galaxies displaced towards brighter
magnitudes and with a narrower dispersion.  The two-wing Gaussian is
parameterized as
\begin{equation}
\label{gaussian_2wing}
\begin{array}{ll}
\phi(M) \;\mathrm{d}M & = \phi_0 e^{-( M_0 - M ) ^2 / 2 \Sigma_a^2}\; \mathrm{d}M \; {\rm for}\; M\le M_0\\
           & = \phi_0 e^{-( M_0 - M ) ^2 / 2 \Sigma_b^2}\; \mathrm{d}M \; {\rm for}\; M\ge M_0\\
\end{array}
\end{equation}
where $M_0$ is the peak magnitude, and $\Sigma_a$ and $\Sigma_b$ are
the dispersion values for the 2 wings. The improvement of the
composite LFs over pure Schechter LFs is most marked for the
early-type LF, because of its bounded behavior at faint magnitudes
which is better adjusted by a Gaussian than a Schechter function, and
for the late-type LF which cannot be fitted by a single Schechter
function at simultaneously faint and intermediate magnitudes.

The ESS $R_\mathrm{c}$ LFs are measured for both the
$R_\mathrm{c}\le20.5$ and $R_\mathrm{c}\le21.5$ samples in Paper~I.
Here we choose to use the LFs measured from the deeper sample because
the faint-end of the LF is better defined than from the shallower
sample.  Composite fits to the LFs for the $V\le21.0$ and $B\le22.0$
samples have not been performed in Paper~I. Here, we however need
these fits in order to use the constraints on galaxy evolution
provided by the $V$ and $B$ faint number counts. Instead of performing
the STY composite fits in the $V$ and $B$ bands, which requires
particular care because of the number of parameters involved, we
prefer to use the simpler approach of converting the $R_\mathrm{c}$ LF
parameters into the $V$ and $B$ bands. We showed in Paper~I that the
shift in magnitude of the $V$ and $B$ LFs with respect to the
$R_\mathrm{c}$ is close to the mean $V-R_\mathrm{c}$, $B-R_\mathrm{c}$
\resp~color of the galaxies in the considered spectral
class. Moreover, the selection biases affecting the $V$ and $B$
samples (a deficiency in galaxies bluer than $B-R_\mathrm{c}\simeq1.5$
and $V-R_\mathrm{c} \simeq0.5$ \resp, due to the selection of the
spectroscopic sample in the $R_\mathrm{c}$ band) affect the faint-end
of the late-type LF when fitted by pure Schechter functions, making it
flatter than in the $R_\mathrm{c}$ band. Conversion of the composite
$R_\mathrm{c}$ LF into the $B$ and $V$ bands allow us to circumvent
the problem of incompleteness in the $B$ and $V$ bands.

Figure~\ref{lf_local_RVB} plots the LFs for the 3 galaxy types in the
$R_\mathrm{c}\le21.5$, $V\le21.0$ and $B\le22.0$ samples.  The points
represent the SWML solutions derived with
$(\Omega_m,\Omega_\Lambda$)=(1.0,0.0) in Paper~I. For the
$R_\mathrm{c}\le21.5$ sample (left panels), the curves show the
composite fits derived in Paper~I; note that for the late-type LF, we
have adopted the intermediate slope $\alpha=-0.3$ (between the values
$\alpha=-0.8$ and $\alpha=0.39$ measured from the
$R_\mathrm{c}\le20.5$ and $R_\mathrm{c}\le21.5$ samples; a slope
$\alpha\simeq-0.3$ is also measured for Sm/Im galaxies in the Virgo
cluster \citep{jerjen97b}).  The parameters for the $V$ and $B$
composite LFs (middle and right panels \resp~of
\fg~\ref{lf_local_RVB}) are then derived from those for the
$R_\mathrm{c}\le21.5$ sample by applying the mean
$M(V)-M(R_\mathrm{c})$ and $M(B)-M(R_\mathrm{c})$ colors for each
spectral class (we use the $R_\mathrm{c}\le20.5$ sample for the color
estimation, rather than the $R_\mathrm{c}\le21.5$ sample, as the
completeness at magnitudes fainter $R_\mathrm{c}=20.5$ is biased in
favor of red objects). The colors are those listed in Table 4 of
Paper~I: $M(V)-M(R_\mathrm{c})=0.5, 0.4, 0.4$ and
$M(V)-M(R_\mathrm{c})=1.4, 1.0, 0.8$ for early-type,
intermediate-type, and late-type galaxies respectively.

Note that the SWML points in \fg~\ref{lf_local_RVB} account for the
incompleteness per apparent magnitude interval, as described by
\citet{zucca94}. For the SWML points, a bin size of $\Delta M =
0.48$\mag~is used in all filters (smaller or larger bin sizes within a
factor 2 yield similar curves). As the amplitudes of the composite
fits and the SWML solutions in the \fg~\ref{lf_local_RVB}, are so far
undetermined (they are measured later on in \scts~\ref{phistar} and
\ref{counts}), we adopt the following: we use the same normalization
of the SWML curves as used in Paper~I (see Table 3), using the
amplitude $\phi^*$ measured from the pure Schechter fits; then for
each sample, the composite is adjusted by least-square fit to the SWML
points (with the ratio $\phi_0/0.4\ln10\;\phi^*$ between the Gaussian
and Schechter component kept fixed to the values in
Table~\ref{lf_BVR}).

Figure~\ref{lf_local_RVB} shows that the composite LFs provide good
adjustment to the SWML solutions for each of the 3 spectral classes in
each filter.  In particular, the simple color shift used to define the
composite LFs in the $V$ and $B$ bands provides good adjustments to
the SWML points in both bands, despite the color biases affecting the
redshift completeness of these samples. Note that the composite
spectral-type LFs derived from the $R_\mathrm{c}\le21.5$ sample
provide satisfying adjustment to the SWML points for the
$R_\mathrm{c}\le20.5$ spectral-type samples.

In Paper~I, the LFs are measured for cosmological parameters
$\Omega_m=1.0$ and $\Omega_\Lambda=0.0$. Because the faint number
counts which we use below to constrain the ESS evolution rate are
sensitive to the cosmological parameters, we have converted these
values to $\Omega_m=0.3$ and $\Omega_\Lambda=0.7$, the currently
favored parameters
\citep{riess98,perlmutter99,phillips01,tonry03}. Again, rather than
re-running the composite fits, we apply the empirical corrections
derived by \citet{lapparent-lc}, as follows.  When changing from
($\Omega_m,\Omega_\Lambda$)=(0.3,0.7) to
($\Omega_m,\Omega_\Lambda$)=(1.0,0.0), the variation in absolute
magnitude due to the change in luminosity distance is $\Delta
M\simeq0.3 ^\mathrm{mag}$ at $z\simeq0.3$, the peak redshift of the
ESS (see \fgs~\ref{nz_R}-\ref{nz_B}).  This empirical correction is
confirmed by the results from \citet{fried01} and \citet{blanton01},
who calculate galaxy LFs in both cosmologies. \citet{lapparent-lc}
also apply a correction to the Schechter parameter $\alpha$, with
$\delta\alpha\sim\Delta M/3$, due to the strong correlation between
the $M^*$ and $\alpha$ parameters in a Schechter
parameterization. Here we neglect this correction, which would amount
to $\delta\alpha\sim0.1$ for a pure Schechter parameterization; this
value is comparable or smaller than the 1-$\sigma$ uncertainty in the
faint-end slope $\alpha$ of the Schechter component for the ESS
intermediate-type and late-type LFs, and than the 1-$\sigma$
uncertainty in the dispersion $\Sigma_a$ and $\Sigma_b$ of the 2-wing
Gaussian fitted to the ESS early-type LF (see Table~\ref{lf_BVR}).

Table~\ref{lf_BVR} lists the resulting LF shape parameters for the
Gaussian and Schechter components ($M_0$, $\Sigma_a$, $\Sigma_b$,
$M^*$, $\alpha$; for a symmetric Gaussian, $\Sigma$ is listed in the
$\Sigma_a$ column): conversion to
($\Omega_m$,$\Omega_\Lambda$)=(0.3,0.7) is obtained by shifting all
values of $M_0$ and $M^*$ by $-0.3 ^\mathrm{mag}$.  The last column of
Table~\ref{lf_BVR} lists the ratio of amplitude
$\phi_0/0.4\ln10\;\phi^*$ between the Gaussian and Schechter component
derived from the composite fits to the $R_\mathrm{c}\le21.5$ sample in
Paper~I. We adopt the same values of this ratio for the $V$ and $B$
LFs.

\section{The amplitude of the ESS luminosity functions \label{phistar} } 

The amplitude of the LF is proportional to the mean density of
galaxies brighter than some absolute magnitude threshold. It is
therefore a useful indicator of the large-scale variations in the
luminous component of the matter density in the Universe, and its
possible evolution with redshift.

\begin{figure}
\resizebox{\hsize}{!}
{\includegraphics{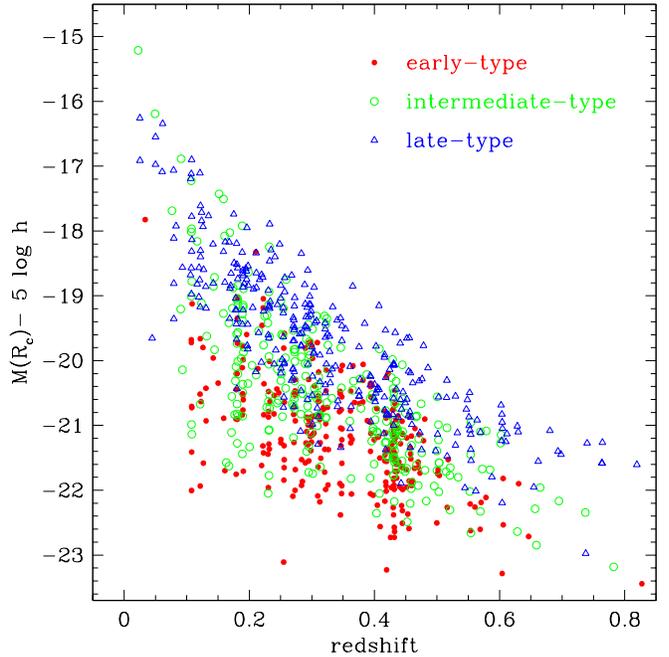}}
\caption{Absolute $R_\mathrm{c}$ magnitudes as a function of redshift
  $z$ for the ESS galaxies with $R_\mathrm{c}\le21.5$, using $H_0 =
  100 h$ km s$^{-1}$ Mpc$^{-1}$, $\Omega_m=0.3$, and
  $\Omega_\Lambda=0.7$. This graph shows how the apparent limiting
  magnitude biases the range of absolute magnitudes detected at
  increasing redshift, and how the varying K-corrections per spectral
  type affect the faintest absolute magnitude reached at a given
  redshift.}
\label{abs_z_R}
\end{figure}

\begin{figure}
\resizebox{\hsize}{!}
{\includegraphics{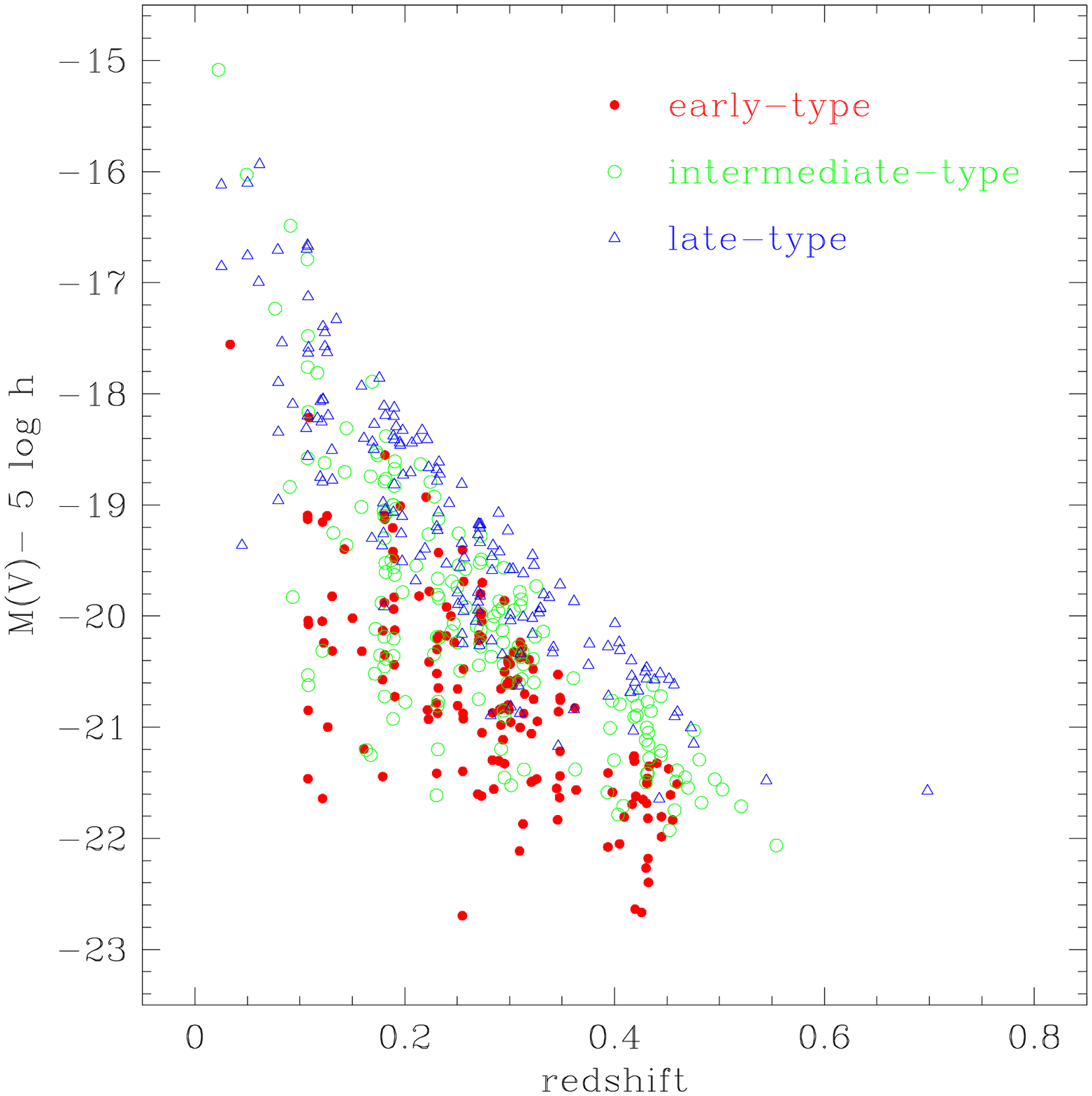}}
\caption{Same as \fg~\ref{abs_z_R} for $V\le21.0$.}
\label{abs_z_V}
\end{figure}

\begin{figure}
\resizebox{\hsize}{!}
{\includegraphics{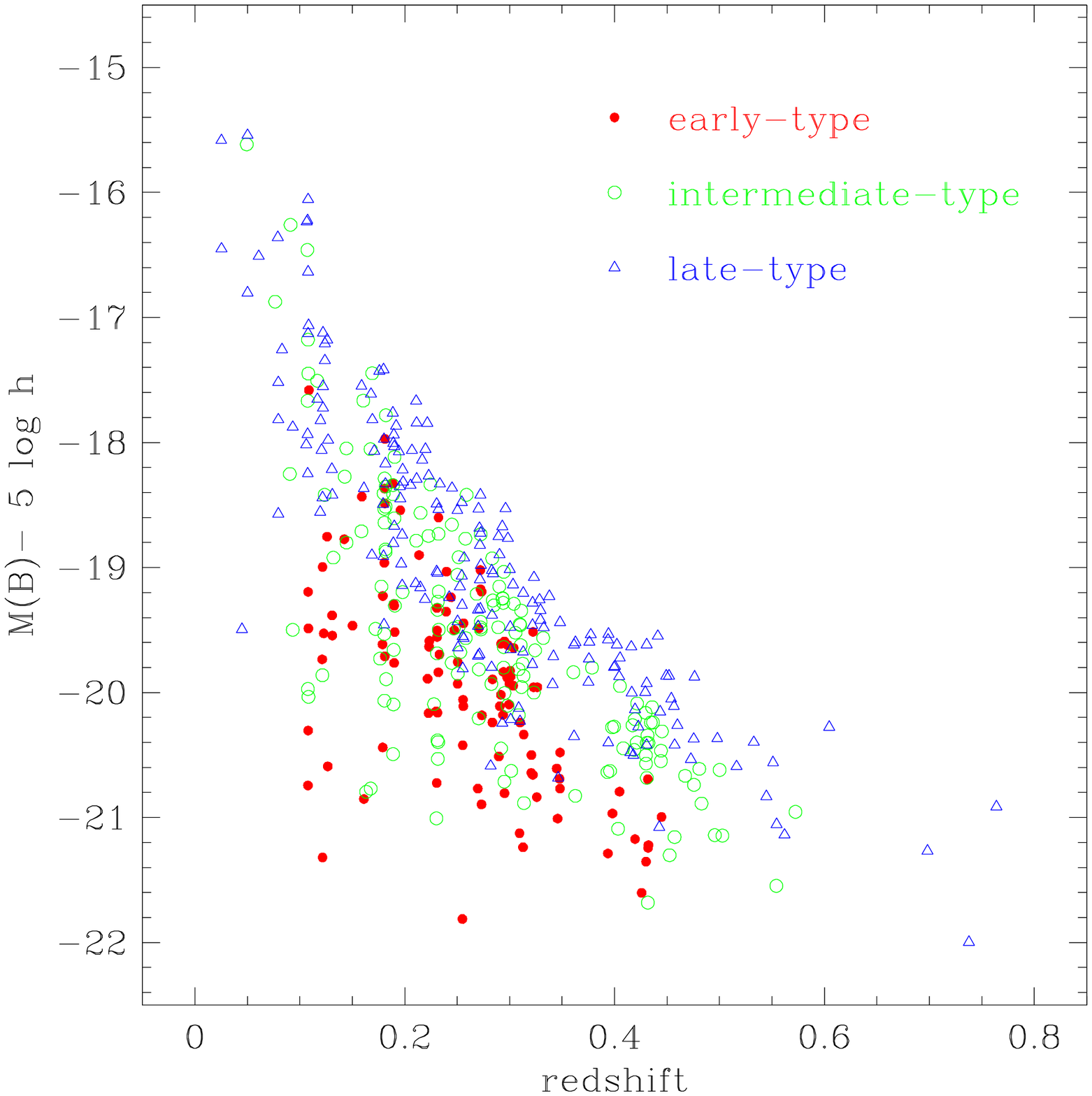}}
\caption{Same as \fg~\ref{abs_z_R} for $B\le22.0$.}
\label{abs_z_B}
\end{figure}

Several other redshift surveys probing the same redshift range as the
ESS survey ($0.1\la z\la0.5$) have detected signs of evolution in the
intrinsic galaxy LFs (see \sct~\ref{evol_comp} below).  The major
difficulty in measuring evolution of the LF with redshift originates
from the limit in apparent magnitude which affects most redshift
surveys. The flux limit results in the detection of galaxies in an
absolute magnitude range which narrows with increasing redshift to the
brightest galaxies. Figure~\ref{abs_z_R} shows how the absolute
magnitudes of the faintest detected ESS galaxies is a function of
redshift and spectral-type (via the K-correction): the limiting curve
at faint magnitudes is defined by replacing in \eq~\ref{abs_mag} the
apparent magnitude $m$ with the $R_\mathrm{c}=21.5$ magnitude limit,
and the K-correction $K(T_\mathrm{S},z)$ with the smallest value over
the considered spectral class at each $z$. Faint galaxies ($-18+5\log
h \la M_(R_\mathrm{c}) \la -15+5\log h$) are exclusively detected at
$z \la 0.2$ in the ESS, whereas only bright early-type and
intermediate-type galaxies with $M(R_\mathrm{c})\le-21.0+5\log h$, and
bright late-type galaxies with $M(R_\mathrm{c})\le-20.0+5\log h$ are
detected at $z\ge0.5$.  In the full ESS redshift range $0.1\la
z\la0.5$, only galaxies in the magnitude interval $-22.0+5\log h \la
M_{R_\mathrm{c}} \la -21.0+5\log h$ can be observed, which is clearly
too narrow for deriving any constraint on the evolution in the shape
of the LF. Figures~\ref{abs_z_V} and \ref{abs_z_B}, which show the
absolute $V$ and $B$ magnitudes versus redshift for the 3 spectral
classes in the $V\le21.0$ and $B\le22.0$ samples \resp, display
similar effects. \citet{driver01} showed that these biases strongly
affect the usual tests for galaxy evolution based on the \emph{shape}
of the LF.

The other limitation for detecting redshift evolution in the ESS is
the limited statistics: separation of each of the 3 spectral classes
into even as few as 2 redshifts intervals would yield large
uncertainties in the measured shape of the LFs, which would make
insignificant any reasonable difference between the high and low
redshift LFs. For the ESS, we can only examine whether the ``general''
LF, \ie the LF summed over all ESS spectral types, evolves with
redshift. Here we consider the LF at $R_\mathrm{c}\le20.5$, as varying
incompleteness at fainter magnitudes may act as evolution. For
$(\Omega_m,\Omega_\Lambda)=(1.0,0.0)$, the ESS general LF for
$R_\mathrm{c}\le20.5$ can be fitted by a Schechter function with
$M^*=-20.95\pm0.12+5\log h$ and $\alpha=-1.15\pm0.09$ (STY fit).  The
corresponding LF in the redshift interval $z<0.3$ at
$R_\mathrm{c}\le20.5$ has $M^*=-21.17\pm0.21+5\log h$ and
$\alpha=-1.18\pm0.11$; at $z>0.3$, $M^*=-21.17\pm0.28+5\log h$ and
$\alpha=-1.63\pm0.31$. The values of $M^*$ are identical in the 2
redshift ranges, and the slope $\alpha$ differs by 1.4-sigma; using
($\Omega_m,\Omega_\Lambda$)=(0.3,0.7) yields similar
conclusions. Given the difficulty in determining the parameter
$\alpha$ in a Schechter fit (Paper~I), these results are consistent
with the hypothesis of null evolution in the shape of the ESS
$R_\mathrm{c}$ ``general'' LF in the redshift range $0.1\la z\la0.5$.
This is however no proof that the individual spectral-type LFs do not
vary in shape: one could imagine redshift variations in the shape of
the intrinsic LFs which would conspire to combine into a constant
``general'' LF.

\citet{driver01} did suggest that reliable constraints on evolution of
the \emph{shape} of the galaxy LF at $z\la1.25$ may only be derived
from samples as deep as the Hubble Deep Field \citep{williams96},
which reaches $z\sim3.5$ \citep{sawicki97}; \citet{driver01}
recommends that such analyses be rather based on the bi-variate
brightness distribution for galaxies (the function which describes the
galaxy bi-variate distribution in absolute magnitude and mean surface
brightness). To circumvent the difficulty in measuring evolution in
the shape of the LF, we assume in the following that the \emph{shape}
of the intrinsic LFs for the ESS early-type, intermediate-type, and
late-type galaxies (as listed in Table~\ref {lf_BVR}) is \emph{not}
evolving with redshift. Any possible evolution will then be detected
as evolution in the amplitude of the LF.

\subsection{Measuring the amplitude of the luminosity function \label{phistar_equat} } 

Once the shape of the LF is determined, its amplitude can be
determined in a second stage. We separate the amplitude $\Phi$ from
the ``shape'' component $\varphi(M)\;\mathrm{d}M$:
\begin{equation}
\begin{array}{ll}
\label{schechter2}
\phi(M) \;\mathrm{d}M = & \Phi\;\varphi(M)\;\mathrm{d}M.\\
\end{array}
\end{equation}
For a Gaussian LF, $\Phi=\phi_0$, and for a pure Schechter LF,
$\Phi=0.4\ln10\;\phi^*$ (see \eqs~\ref{gaussian} and
\ref{schechter_lum}).  In a survey where the detected brightest and
faintest absolute magnitudes are $M_1$ and $M_2$ \resp, one can
calculate the mean density of galaxies with $M_1\le M\le M_2$, denoted
$\overline{n}\;(M_1\le M\le M_2)$, which is related to $\Phi$ by
\begin{equation}
\label{phi_star}
\Phi = \frac{\overline{n}\;(M_1\le M\le M_2)}{\int_{M_1}^{M_2}\varphi(M) \;\mathrm{d}M};
\end{equation}

In a magnitude-limited survey, the absolute magnitudes
$M_\mathrm{bright}(z)$ and $M_\mathrm{faint}(z)$ of the detected
brightest and faintest galaxy \resp~vary with redshift (see
\fgs~\ref{abs_z_R}-\ref{abs_z_B}). Calculation of the mean
density $\overline{n}\;(M_1\le M\le M_2)$ therefore requires to
correct the observed number of galaxies to the expected number if the
survey was limited to the constant absolute magnitude interval $M_1\le
M\le M_2$. The correction is obtained by multiplying the observed
number of galaxies at redshift $z$ by the inverse of the selection
function $S(z)$ defined as
\begin{equation}
S(z) = \frac{\int_{\max(M_\mathrm{bright}(z),M_1)}^{\min(M_\mathrm{faint}(z),M_2)}
\varphi(M) \;\mathrm{d}M}{\int_{M_1}^{M_2}\varphi(M)\;\mathrm{d}M};
\label{selfunc}
\end{equation}
For the ESS, we take in all filters $M_1=-30+5\log h$, and
$M_2=-16+5\log h$ (see \fgs~\ref{abs_z_R}-\ref{abs_z_B}). $S(z)$ then
measures the fraction of galaxies with $M_1\le M\le M_2$ at redshift
$z$ which are included in the survey.

Following \citet{davis82b}, we define 3 estimators for the mean
density $\overline{n}\;(M_1\le M\le M_2)$ which are unbiased by the
apparent magnitude limit of the survey (in the following, although we
omit to mention $M_1\le M\le M_2$, all quoted densities refer to that
interval). If $N(z)\;\mathrm{d}z$ is the observed number of galaxies in a shell
$\mathrm{d}z$ at redshift $z$, $N(z)\;\mathrm{d}z/S(z)$ is the expected number of galaxies
with $M_1\le M\le M_2$. A first estimator of the mean density is defined
by \citet{davis82b} as
\begin{equation}
\label{n1}
n_1(z_\mathrm{max}) = \frac{1}{V(z_\mathrm{min},z_\mathrm{max})}
\int_{z_\mathrm{min}}^{z_\mathrm{max}}\frac{N(z)}{S(z)}\;\mathrm{d}z,
\end{equation}
where $z_\mathrm{min}$ and $z_\mathrm{max}$ are arbitrary choices of the smallest
and largest redshift over which the integrals are performed, and
$V(z_\mathrm{min},z_\mathrm{max})$ is the total volume of the survey between these
limits:
\begin{equation}
\label{vol}
V(z_\mathrm{min},z_\mathrm{max}) = \int_{z_\mathrm{min}}^{z_\mathrm{max}}\;\frac{\mathrm{d}V}{\mathrm{d}z}\;\mathrm{d}z
\end{equation}
($\mathrm{d}V$ is the comoving volume element at redshift $z$,
see \eq~\ref{dv}). Here we use $z_\mathrm{min}=0.1$ (see
\sct~\ref{evol}).  Note that because $1/S(z)$ rises sharply with
redshift, the $n_1$ estimator heavily weights distant
structures. \citet{davis82b} also showed that $n_1$ is close to the
minimum variance estimator of the mean density.

\citet{davis82b} define another estimator by equating the
observed number of galaxies with the expected number in a homogeneous
universe:
\begin{equation}
\label{n3}
n_3(z_\mathrm{max}) = \frac{\int_{z_\mathrm{min}}^{z_\mathrm{max}}N(z)\;\mathrm{d}z}{\int_{z_\mathrm{min}}^{z_\mathrm{max}}S(z)\;\frac{\mathrm{d}V}{\mathrm{d}z}\;\mathrm{d}z}.
\end{equation}
This estimator is much more stable than $n_1$, as all observed
galaxies are weighted equally. It however heavily weights
galaxies near the peak of the redshift distribution \citep{davis82b}.

Finally, \citet{davis82b} define a third and intermediate estimator by
averaging the expected density across radial shells:
\begin{equation}
\label{n2}
n_2(z_\mathrm{max}) = \frac{1}{z_\mathrm{max}-z_\mathrm{min}}\int_{z_\mathrm{min}}^{z_\mathrm{max}}\frac{N(z)}{S(z)\;\frac{\mathrm{d}V}{\mathrm{d}z}}\;\mathrm{d}z
\end{equation}
The interest of the $n_2$ estimator is that its differential value 
\begin{equation}
\label{dn2}
\mathrm{d}n_2 =\frac{N(z)}{S(z)\;\frac{\mathrm{d}V}{\mathrm{d}z}}\;\mathrm{d}z
\end{equation}
can also be calculated as a function of redshift, and allows one to
examine the variations of the local galaxy density with redshift.
Note that the 3 estimators $n_1$, $n_2$, and $n_3$ can be calculated
to a varying depth $z_\mathrm{max}$ and are thus defined as functions
of $z_\mathrm{max}$. In a fair sample of the galaxy distribution, that
is a sample which is significantly larger than the largest density
fluctuations, all 3 estimators should converge to a common value as
$z_\mathrm{max}$ reaches the sample depth.

A fourth estimator, which we test here, is that proposed by
\citet{efstathiou88a}.  This estimator is similar to $n_1$ but the
shell $\mathrm{d}z$ is taken to be infinitely small so that galaxies
can be counted one by one, and the integral can be re-written as a sum
over the galaxies in the sample:
\begin{equation}
\label{n_eep}
n_\mathrm{EEP}(z_\mathrm{max}) = \frac{1}{V(z_\mathrm{min},z_\mathrm{max})}\sum_{i=1}^{N_\mathrm{gal}}\frac{1}{S(z_i)},
\end{equation}
where $z_i$ is the redshift of each galaxy, $N_\mathrm{gal}$ galaxies are
observed in the redshift range $z_\mathrm{min}-z_\mathrm{max}$, and
the volume $V(z_\mathrm{min},z_\mathrm{max})$ is defined in
\eq~\ref{vol}.

The variations with $z_\mathrm{max}$ of the estimators $\Phi_1$,
$\Phi_2$, $\Phi_3$, and $\Phi_\mathrm{EEP}$, based on
$n_1(z_\mathrm{max})$, $n_2(z_\mathrm{max})$, $n_3(z_\mathrm{max})$,
and $n_\mathrm{EEP}(z_\mathrm{max})$ respectively, can then be defined
using \eq~\ref{phi_star}.  In practise, the integrals in the $n_1$,
$n_2$ and $n_3$ estimators are calculated as discrete sums over a
finite redshift bin $\Delta z$, and the selection function correction
is approximated as $S(z_\mathrm{c})$ where $z_\mathrm{c}$ is the
central redshift of the bin. However, because the second derivative of
the selection function is positive, this yields an underestimate of
the expected number of galaxies. The resulting systematic error in
$\Phi$ is $\la 1$\%.  It is reduced to $\sim0.1$\% by replacing
$S(z_\mathrm{c})$ by its average value over the bin, which we adopt
for the $n1$, $n_2$ and $n_3$ estimators.  The $n_\mathrm{EEP}$
estimator is a priori unbiased by this effect, as galaxies are
considered one by one, and $S(z)$ is calculated at the redshift of
each galaxy. Our tests with mock ESS catalogues (described in the next
\sct) however show that the $n_\mathrm{EEP}$ estimator tends to
over-estimate the true $\Phi$ by $\sim1$\% for uniform distributions
with $\sim200$ points; this bias disappears for distributions with
more than $\sim2000$ points. In the ESS spectral classes for which
$N_\mathrm{gal}\sim100-300$ galaxies, a $\sim1$\% bias is small
compared to the random uncertainties in the $n1$, $n_2$ and $n_3$
estimators (see next \sct).

\subsection{Estimation of errors                                 \label{errors} } 

We estimate the random errors in the amplitude of the LFs for the ESS
by generating mock ESS distributions with $\sim 240$, $\sim2400$ and
$\sim24\,000$ points, and a Schechter LF defined by $M^*=-19.2$ and
$\alpha=-1.4$ (using composite LFs as those listed in
Table~\ref{lf_BVR} would not change any of the reported results); the
measured values of $\phi^*$ are then in proportion of the total number
of galaxies in each simulation.  The mock distributions have no
built-in clustering, and are thus uniform spatial distributions
modulated by the selection function (see \eq~\ref{selfunc}). To
nevertheless take into account the effects of \emph{large-scale}
clustering, we also introduce in the simulations density fluctuations
in redshift which mimic large-scale structure: fluctuations resembling
those measured in the ESS (measured as the departure from the redshift
distribution for a uniform distribution), and various transformations
of these including the inverse of the ESS fluctuations. We neglect
density fluctuations transverse to the line-of-sight, as they are
expected to have a negligeable effect on the calculation of the mean
density.

For each type of simulation (choice of $\phi^*$ and fluctuations in
density), we generate 100 samples with different seeds for the random
generators. The random errors in the various estimators of $\phi^*$
are then calculated directly from the variance among the 100
realizations of the same distribution, and are functions of the
redshift $z_\mathrm{max}$ out to which \eq~\ref{n1}--\ref{n2}~are
integrated.  We measure that the variance in the estimator $\Phi_3$
of $\phi^*$ (based on $n_3$) varies as
\begin{equation}
\label{sigma_n3}
\frac{\sigma(\Phi_3)}{\Phi_3} \simeq \frac{1.0}{\sqrt{N_\mathrm{gal}(z_\mathrm{min},z_\mathrm{max})}}
\end{equation}
where $N_\mathrm{gal}(z_\mathrm{min},z_\mathrm{max})$ is the number of
galaxies in the survey which lie in the redshift interval
$[z_\mathrm{min},z_\mathrm{max}]$.  This is valid out to the most
distant galaxy included in the survey.  For the $n_2$ estimator of
$\phi^*$, we measure
\begin{equation}
\label{sigma_n2}
\frac{\sigma(\Phi_2)}{\Phi_2} \simeq \frac{1.5}{\sqrt{N_\mathrm{gal}(z_\mathrm{min},z_\mathrm{max})}};
\end{equation}
This is also valid out to the redshift where $N_\mathrm{gal}$
corresponds $\sim95$\% of the total number of galaxies in the sample
($z\simeq0.6$ in the ESS). For the $n_1$ and $n_\mathrm{EEP}$
estimators, the relative variance in $\phi^*(z_\mathrm{zmax})$ rise
slowly from $1.0/\sqrt{N_\mathrm{gal}(z_\mathrm{min},z_\mathrm{max})}$
at low redshift to
$1.5/\sqrt{N_\mathrm{gal}(z_\mathrm{min},z_\mathrm{max})}$ when
$N_\mathrm{gal}$ reaches $\sim85$\% of the total number of galaxies in
the sample ($z\simeq0.4$ in the ESS) and can grow to
$2.0/\sqrt{N_\mathrm{gal}(z_\mathrm{min},z_\mathrm{max})}$ at larger
distances (in the redshift interval $0.4<z<0.6$ for the ESS).

\citet{davis82b} calculate that for the minimum variance estimate of
the mean density (close to $n_1$), the relative uncertainty due to
galaxy clustering within the finite volume sample is
\begin{equation}
\label{dens_error}
\frac{\sigma(\overline{n})}{\overline{n}} \sim
\sqrt{\frac{J_3(z_\mathrm{min},z_\mathrm{max})}{V(z_\mathrm{min},z_\mathrm{max})}},
\end{equation}
where $4\pi J_3$ is the volume integral of the 2-point galaxy
correlation in the survey volume bounded in redshift by
$[z_\mathrm{min},z_\mathrm{max}]$. By using the correlation function
$\xi(r_\mathrm{P},\pi)$ measured from the ESS by \citet{slezak04}, we
calculate the values of $J_3(z_\mathrm{min},z_\mathrm{max})$, and we
confirm that the relative uncertainties derived from
\eq~\ref{dens_error} for one spectral class are comparable to those
given by \eq~\ref{sigma_n3} for $z_\mathrm{min}=0.1$ and $0.1\le
z_\mathrm{max}\le0.6$, with \eq~\ref{dens_error} becoming slightly
larger than \eq~\ref{sigma_n3} at $z\la0.3$.

As pointed by \citet{lin96}, \eq~\ref{dens_error} and the errors
calculated from the mock distributions {\em underestimate} the error
in $\phi^*$, as they do not take into account the uncertainty in the
shape of the LF: indeed, in our mock distributions, the LF is not
re-measured, as we assume that the true value is known. In that case
(when the {\it input} values of $M^*$ and $\alpha$ are used to
calculate $\phi^*$), all 4 density estimators recover the true value
to $0.1-1$\% (see \eqs~\ref{sigma_n3} and \ref{sigma_n2}). The
systematic underestimation of $\phi^*$ by 20\% calculated by
\citet{willmer97} for LFs derived by the STY estimator, could be a
measure of this effect, if one assumes that $\phi^*$ was measured
using the biased values of $M^*$ and $\alpha$ (but the author
unfortunately does not specify).  Here, however, we do not need to
evaluate this additional source of uncertainty because the large
amplitude of the density fluctuations on scales of 100\hmpc~ which are
present in the ESS survey cause systematic variations in the
determination of both $\phi^*$ and $\phi_0$ (see \sct~\ref{evol})
which dominate all the other sources of error discussed here.

\subsection{Evolution in the ESS mean density \label{evol} }

\begin{figure}
  \resizebox{\hsize}{!}
  {\includegraphics{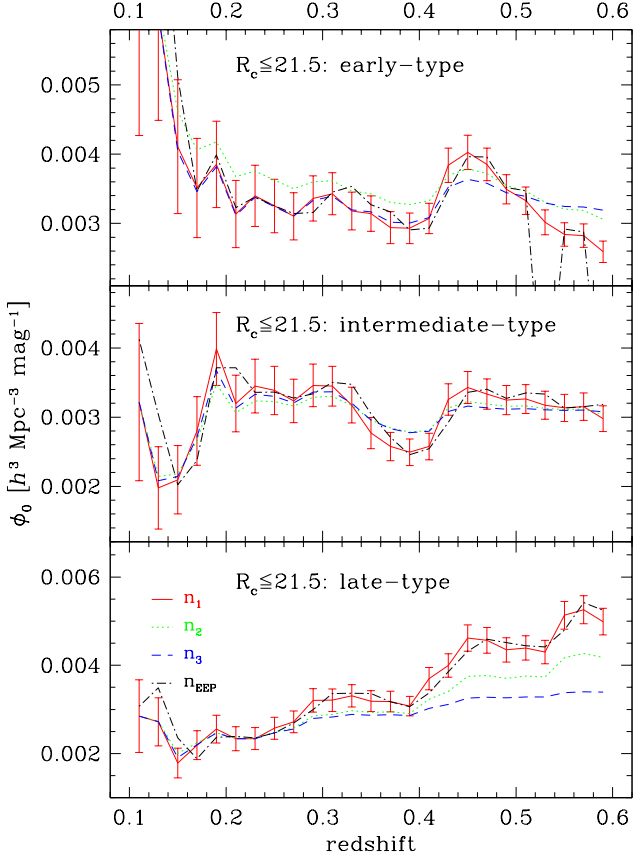}}
\caption{The amplitude $\phi_0$ of the Gaussian component of the
  composite luminosity functions measured for the ESO-Sculptor
  $R_\mathrm{c}\le21.5$ sample, using the estimators $n_1$ (red solid
  line), $n_2$ (green dotted line), $n_3$ (blue dashed line), and
  $n_\mathrm{EEP}$ (black dot-dashed line) defined in
  \eqs~\ref{n1}--\ref{n_eep}, as a function of redshift (denoted
  $z_\mathrm{max}$ in the text).  Each panel corresponds to a
  spectral-type.  For clarity, only the random errors for the $n_1$
  estimator, calculated using \eq~\ref{sigma_n3} are shown (see text
  for details). The error bars for the other curves have similar
  amplitude, except for $n_2$, for which they are 50\% wider.}
\label{phistar_type}
\end{figure} 

\begin{figure}
  \resizebox{\hsize}{!}
    {\includegraphics{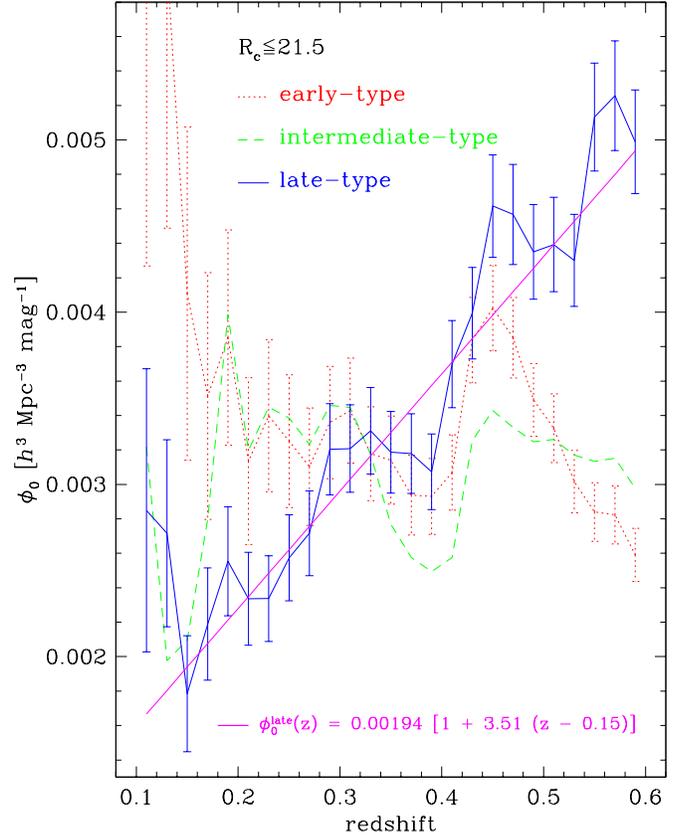}}
\caption{The amplitude $\phi_0$ of the Gaussian component of the
  composite luminosity functions measured for the ESO-Sculptor
  $R_\mathrm{c}\le 21.5$ sample, using the estimator $n_1$ defined in
  \eq~\ref{n1}, as a function of redshift $z_\mathrm{max}$;
  early-type, intermediate-type, late-type galaxies correspond to the
  red dotted, green dashed, and blue solid lines respectively.  For
  clarity, the random errors for the intermediate-type galaxies are
  not shown; they are comparable to those for the early-type and
  late-type galaxies (see \eq~\ref{sigma_n3}). The linear fit of
  \eq~\ref{phi_evol_ratio} is also plotted (magenta solid straight
  line).}
\label{phistar_n1}
\end{figure}

\begin{figure}
\resizebox{\hsize}{!}{\includegraphics{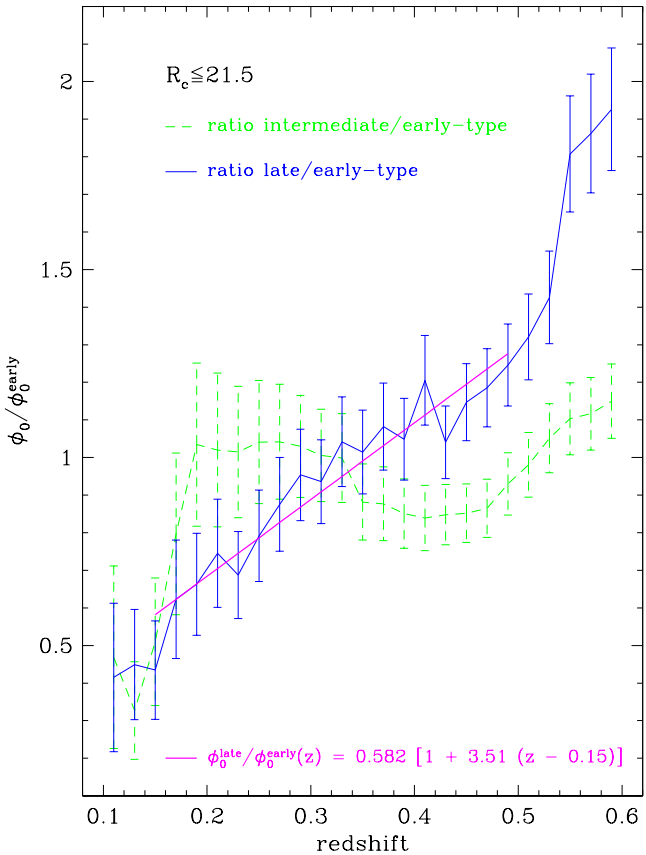}}
\caption{The amplitude $\phi_0(z_\mathrm{max})$ of the Gaussian
  component of the composite luminosity functions measured from the
  ESO-Sculptor $R_\mathrm{c}\le 21.5$ sample for the intermediate-type
  (green dashed line) and late-type galaxies (blue solid line), both
  normalized to $\phi_0(z_\mathrm{max})$ for the early-type
  galaxies. All values of $\phi_0$ are derived using the $n_1$
  estimator defined in \eq~\ref{n1}. The linear fit to the relative
  late-type density given in \eq~\ref{phi_evol_late} is also shown
  (magenta solid straight line).}
\label{phistar_ratio}
\end{figure}

\begin{figure}
\resizebox{\hsize}{!}{\includegraphics{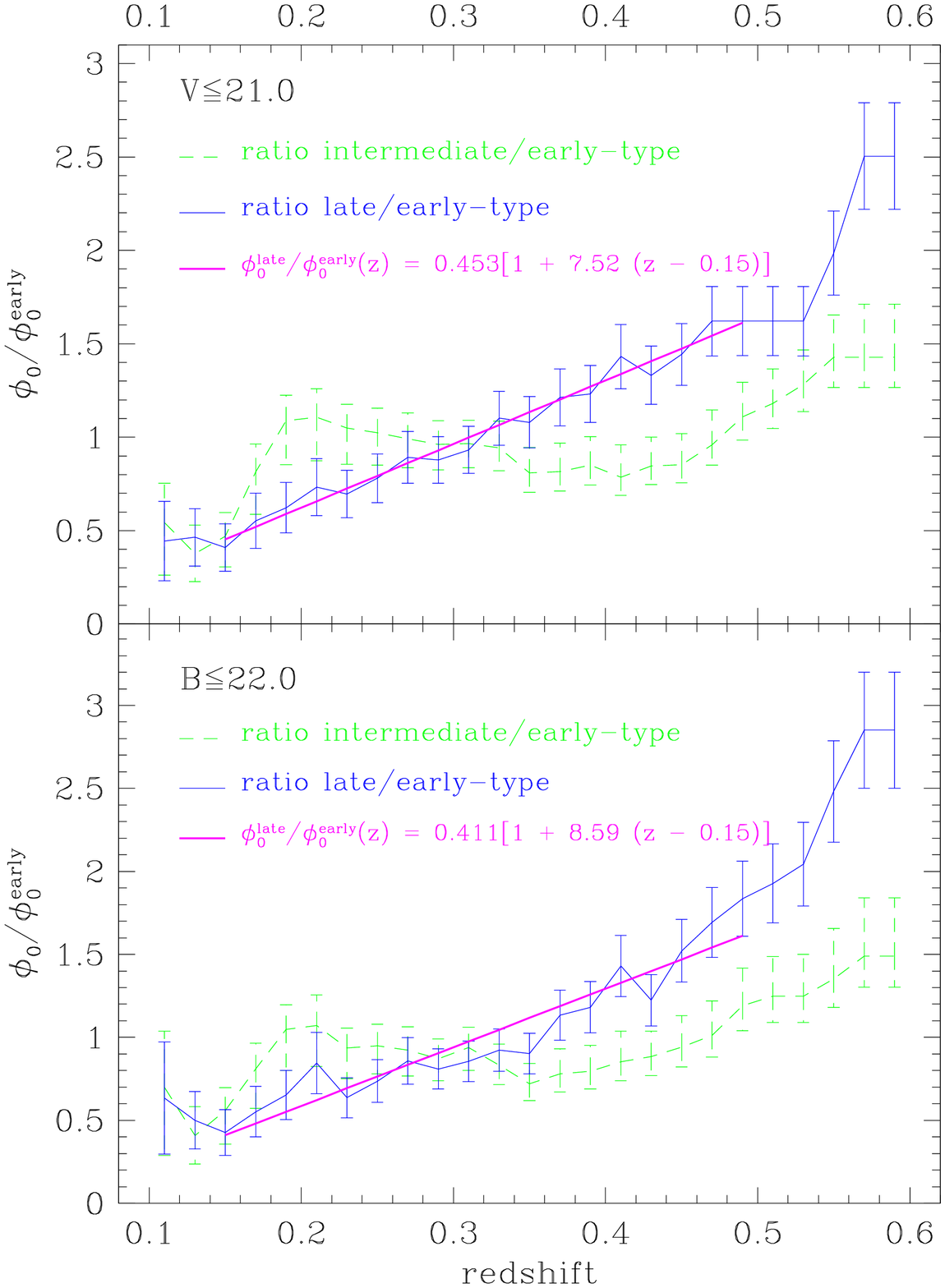}}
\caption{The amplitude $\phi_0(z_\mathrm{max})$ of the Gaussian
  component of the composite luminosity functions measured from the
  ESO-Sculptor $V\le21.0$ sample (top panel) and the $B\le22.0$ sample
  (bottom panel) for the intermediate-type (green dashed line) and
  late-type galaxies (blue solid line), both normalized to
  $\phi_0(z_\mathrm{max})$ for the early-type galaxies. All values of
  $\phi_0$ are derived using the $n_1$ estimator defined in
  \eq~\ref{n1}. The linear fits to the relative late-type density and
  the corresponding parameterization are also shown in each panel
  (magenta solid straight line).}
\label{phistar_VB}
\end{figure}

Figure~\ref{phistar_type} shows for each of the 3 ESS spectral classes
the variations with redshift $z_\mathrm{max}$ of the estimates
$\Phi_1$, $\Phi_2$, $\Phi_3$, and $\Phi_\mathrm{EEP}$ of the amplitude
$\phi_0$ for the Gaussian component of the composite LFs (see
Table~\ref{lf_BVR} and \eqs~\ref{gaussian} and \ref{gaussian_2wing});
these estimates are based on the $n_1(z_\mathrm{max})$,
$n_2(z_\mathrm{max})$, $n_3(z_\mathrm{max})$,
$n_\mathrm{EEP}(z_\mathrm{max})$ estimators of the mean density
\resp~(see \eqs~\ref{n1}--\ref{n_eep}) for the ESS
$R_\mathrm{c}\le21.5$ sample.  All quantities plotted in
\fg~\ref{phistar_type} (and the other figures in this section) use
$z_\mathrm{min}=0.1$, in order to avoid biasing of the density
estimates, as the small volume probed at $z \le 0.1$ suffers from
under-sampling (see \fgs~\ref{abs_z_R}-\ref{abs_z_B}). The estimators
$n_1$, $n_2$, $n_3$ are calculated in steps of $\Delta z=0.02$. The
random errors due to the finite sample size for the $n_1$, $n_3$, and
$n_\mathrm{EEP}$ estimators are evaluated using \eq~\ref{sigma_n3};
for the $n_2$ estimator, \eq~\ref{sigma_n2} is used.

We emphasize that under the assumption of a non-evolving shape of the
intrinsic LFs $\varphi(M)$ for the various galaxy types (see
\sct~\ref{phistar}), examining the variations in $\phi_0$ with
redshift is equivalent to examining the variations in the ESS mean
density with redshift (see \eq~\ref{phi_star}). Moreover, in the
following, we only show and discuss the variations in the amplitude
$\phi_0$ of the Gaussian component of each spectral-type
LF. Nevertheless, because the variations in the amplitude $\phi^*$ of
the Schechter components for the intermediate-type and late-type LFs
are simply proportional to those in $\phi_0$ using the ratio
$\phi_0/0.4 \ln10\,\phi^*$, whose values are listed in
Table~\ref{lf_BVR}, all comments on the variations of $\phi_0$ also
apply to those in $\phi^*$. In order to refer to both $\phi_0$ and
$\phi^*$, we use in the following the generic amplitude $\Phi$.

We first compare the performances of the 4 estimators $n_1$,
$n_2$, $n_3$ and $n_\mathrm{EEP}$.  Figure~\ref{phistar_type} shows
that the $n_\mathrm{EEP}$ estimator \citep{efstathiou88a} yields
nearly indistinguishable results from the $n_1$ estimator. For all
galaxy types, the $n_3$ estimator converges at $z\ga 0.5$ to a value
determined by the ratio of total number of galaxies in the considered
sample by the integral of $\varphi(M)$ (see \eq~\ref{schechter2}); the
asymptotic $n_3$ estimator of $\phi_0$ is however dominated by the
observed number of galaxies at $z\la0.3$ (see
\sct~\ref{phistar_equat}).  In contrast, as the $n_1$ estimator
weights different parts of the survey proportionally to their volume,
it allows one to trace the large-scale density fluctuations across the
ESS redshift interval. As expected, the $n_2$ estimator of $\phi_0$
(\eq~\ref{n2}) yields intermediate values between the $n_1$ and $n_3$
estimators for the 3 galaxy types.  In the following, we thus restrict
the discussion to the extreme cases represented by the $n_1$ and $n_3$
estimators.

Figure~\ref{phistar_n1} provides a direct comparison of the
amplitude $\phi_0$ of the Gaussian component of the LFs based on the
$n_1$ estimator (denoted $\phi_1$ hereafter) for the 3 ESS spectral
types. The 3 curves show a relative depression in the range $0.33\la
z\la0.42$ followed by an excess at $z\sim0.45$. Both features
correspond to large-scale fluctuations in the redshift
distribution. Note that for the 3 galaxy types, the fluctuations in
the $\Phi_1$ about its mean value have a typical
amplitude of $0.001$, corresponding to a relative variation of
$\sim30$\%. These systematic errors largely dominate the random errors
in $\Phi_1$ indicated by the vertical error-bars in
\fg~\ref{phistar_n1} (see also \eqs~\ref{sigma_n3} and
\ref{sigma_n2}). The other marked effect in \fg~\ref{phistar_n1} is
the steady increase of $\Phi_1$ for the late-type galaxies compared to
the more stable behavior for the early-type and intermediate-type
galaxies: a factor $\sim 2.5$ increase is measured from $z\simeq0.15$
to $z\simeq0.55$. We interpret this effect as an evolution in $\Phi_1$
for the late-type galaxies and provide further evidence below.  An
increase of $\Phi_3$ for the late-type galaxies is also visible in
\fg~\ref{phistar_type}, but the effect is smaller as $\Phi_3$ applies
a lower weight to the distant structures relative to those nearby.

\begin{table*}
\caption{$n_1$ and $n_3$ estimators of the amplitude $\phi_0$ of the
Gaussian component of the early-type, intermediate-type, and late-type
luminosity functions for the ESO-Sculptor redshift survey, in the
Johnson $B$, $V$ and Cousins $R_\mathrm{c}$ bands.}
\label{phistar_BVR}
\begin{center}
\begin{tabular}{lcccclcc}
\hline 
\hline 
Sample                & \multicolumn{2}{l}{Early-type} & \multicolumn{2}{l}{Intermediate-type}
& \multicolumn{3}{l}{Late-type}   \\ 
                      & $\Phi_1(0.51)$  & $\Phi_3(0.51)$ & $\Phi_1(0.51)$ & $\Phi_3(0.51)$ 
& $\Phi_1(z)$ & $\Phi_1(0.51)$  & $\Phi_3(0.51)$ \\ 
\hline
$R_\mathrm{c}\le21.5$         & $0.00333$    & $0.00339$   & $0.00326$   & $0.00312$ 
& $0.00194[1+3.51(z-0.15)]$  & 0.00439 & 0.00328  \\
$R_\mathrm{c}\le20.5$         & $0.00324$    & $0.00331$   & $0.00351$   & $0.00321$  
& $0.00177[1+4.45(z-0.15)]$  & $0.00462$ & $0.00319$  \\  
$V\le21.0$                    & $0.00322$    & $0.00335$   & $0.00379$   & $0.00314$        
& $0.00141[1+7.52(z-0.15)]$  & $0.00521$ & $0.00309$  \\  
$B\le22.0$                    & $0.00286$    & $0.00363$   & $0.00357$   & $0.00325$          
& $0.00135[1+8.59(z-0.15)]$  & $0.00551$ & $0.00333$  \\  
\hline
\end{tabular}
\smallskip
\end{center}

\begin{list}{}{}
\item[\underline{Notes:}]
\item[-]The listed values of the luminosity function amplitude
$\phi_0$ are in units of \phiunit, and are obtained with $H_0 = 100 h$
km s$^{-1}$ Mpc$^{-1}$, $\Omega_m=0.3$, and $\Omega_\Lambda=0.7$.
\item[-]The corresponding amplitude $\phi^*$ of the Schechter
component for the intermediate-type and late-type galaxies can be
derived using the values of the ratio $\phi_0/0.4\ln10\;\phi^*$ listed
in Table~\ref{lf_BVR}.
\item[-]The $\Phi_1(z)$ parameterization of the evolving
amplitude $\phi_0$ for the late-type galaxies should be used with
caution (especially in the $V$ and $B$ bands); more realistic
parameterizations are obtained in the $R_\mathrm{c}$, $V$ and $B$
bands from the adjustment of the number counts (see
Table~\ref{evol_BVR}, \sct~\ref{counts}).
\end{list}
\end{table*}

The variations of $\Phi_1$ with redshift in \fg~\ref{phistar_n1} thus
result from the combination of the density fluctuations produced by
the pattern of large-scale structure with the possible evolution in
the amplitude of the LF. A survey with a wider angular extent would be
necessary to average out the effect of large-scale structures
perpendicular to the line-of-sight.  Although the different galaxy
types have different clustering properties \citep{loveday95}, which on
small scales are characterized by the morphology-density relationship
\citep{dressler80}, they do trace the same pattern of walls and voids
at large scales \citep{huchra90}. There has been so far no detection
of systematic variations in the proportions of the different galaxy
types on scales larger than $\sim 10$\hmpc~ (and thus outside galaxy
clusters).  We therefore make the hypothesis that the proportions of
the different galaxy types are constant at large scales, and we
eliminate the fluctuations caused by the large-scale structure by
normalizing $\Phi_1$ for the intermediate and late-type galaxies by
$\Phi_1$ for the early-type galaxies. The resulting relative
variations of $\phi_0$ in the $R_\mathrm{c}\le21.5$ sample are shown
in \fg~\ref{phistar_ratio}. The normalization erases most of the
fluctuations in \fg~\ref{phistar_n1}, and only the deviations from a
distribution having similar large-scale clustering as the early-type
galaxies remain. Whereas the relative density for the
intermediate-type galaxies remains within the narrow interval
$\sim0.8-1.0$ in the redshift interval $0.15\la z\la0.5$, the relative
density of late-type galaxies shows a linear increase by a factor of
nearly 2 in this interval.

Under the assumption that the shape $\varphi(M)$ of the intrinsic LFs
for the various galaxy types are non-evolving with redshift (see
\sct~\ref{phistar}), the systematic increase with redshift in the
$\Phi_1$ estimate of $\phi_0$ for the late-type galaxies detected in
\fg \ref{phistar_ratio} can be interpreted as evidence for evolution
in the ESS. A linear regression to the relative late-type density in
the interval $0.15\le z\le0.5$ yields
\begin{equation}
\label{phi_evol_ratio}\frac{\phi_0^\mathrm{late}}{\phi_0^{\rm early}}(z)= 0.582\;[ 1 + 3.51\,(z-0.15) ];
\end{equation}
this fit is over-plotted in \fg~\ref{phistar_ratio}. Note that the
linear fit is performed in the restricted redshift interval $0.15\le
z\le0.5$ because the survey volume is too small at $z\le0.1$ (see
\fgs~\ref{abs_z_R}-\ref{abs_z_B}) and the redshift survey is too
diluted at $z\ge0.5$ to provide a reliable estimate of $\phi_0$. We
now assume that $\phi_0$ for the early-type galaxies does not evolve
with redshift at $z\le0.5$, and that the redshift evolution in
\eq~\ref{phi_evol_ratio} can be fully attributed to the late-type
galaxies. The parameterization of the late-type density evolution
therefore becomes
\begin{equation}
\label{phi_evol_late} \phi_0^\mathrm{late}(z) = \phi_0^\mathrm{late}(0.15)\;[ 1 + P_{0.15}\,(z-0.15) ]
\end{equation}
with
\begin{equation}
\begin{array}{ll}
P_{0.15} &= 3.51\\
\phi_0^\mathrm{late}(0.15)&=0.00194\,h^3\,\mathrm{Mpc}^{-3}\,\mathrm{mag}^{-1}\\
\end{array}
\label{phi_015} 
\end{equation}
The zero-point $\phi_0^\mathrm{late}(0.15)$ is measured by adjusting the
linear fit to the measured values of the $\Phi_1$ estimate of $\phi_0$
at $z=0.51$ for the late-type galaxies, which is listed in
Table~\ref{phistar_BVR}. The linear evolution model of
\eq~\ref{phi_evol_late} is over-plotted in \fg~\ref{phistar_n1}, with
extrapolation to the redshift intervals $0.1\le z\le0.15$ and $0.5\le
z\le0.6$.  This model follows satisfactorily the increasing trend of
the $\Phi_1$ estimator of $\phi_0$ for the late-type galaxies (recall
that the deviations from the model are mostly caused by the
large-scale structures).

We emphasize that the $\Phi_1(z)$ estimator at redshift $z$ is a
cumulative measure over the redshift interval $0.1-z$, which may
therefore underestimate the evolution rate for the late-type galaxies.
However, the contribution at each redshift is proportional to the
volume sampled; as the volume increases with $z^2$, this puts most of
the weight on the structures at $z$, making $\Phi_1$ a good
approximation to an incremental measure of the density variations with
redshift. We also confirm this result by measuring the estimate of
$\phi_0$ using the incremental estimator of the mean density
$\mathrm{d}n_2$, described in \sct~\ref{phistar_equat}
(\eq~\ref{dn2}). The estimator $\mathrm{d}n_2$ yields a similar rate
of increase in $\phi_0$ as the $n_1$ estimator, but its large error
bars prevent any reliable measure.

Other authors have modeled the evolution in the amplitude $\Phi$ of
the LF as a power of $1+z$ \citep{lilly96,heyl97}, as it converts to a
power of cosmic time for ($\Omega_m$,$\Omega_\Lambda$)=(1.0,0.0):
$1+z\propto t^{-2/3}$ \citep{cole92}. Using the adopted values
($\Omega_m$,$\Omega_\Lambda$)=(0.3,0.7), adjustment of $\phi_0^{\rm
late}/\phi_0^{\rm early}$ by a power-law function defined as
\begin{equation}
\label{phi_evol_power} 
\phi_0^\mathrm{late}(z) = \phi_0^\mathrm{late}(0)\;(1 + z)^\gamma,
\end{equation}
over the redshift interval $0.15\le z\le0.5$, deviates from the linear
model by at most $\sim 0.0001$\phiunit, significantly smaller
than the 1-$\sigma$ random errors in the measurement of $\phi_0^{\rm
late}$ (see \fg~\ref{phistar_n1}).  Therefore, in the limited
redshift range of the ESS, both the linear and power-law models
provide good descriptions of the evolution of the relative density of
late-type galaxies at $R_\mathrm{c}\le21.5$. In \sct~\ref{counts}, we
show that when extrapolated to $z\sim1$, both the linear and power-law
models can also adjust the ESS faint $BVR_\mathrm{c}$ number-counts.

For numerical comparison of the $\Phi_1$ and $\Phi_3$ estimators
of $\phi_0$, we list in Table~\ref{phistar_BVR} their values at
$z_\mathrm{max}=0.51$ for each spectral class of the
$R_\mathrm{c}\le21.5$, $R_\mathrm{c}\le20.5$, $V\le21.0$ and $B\le22.0$
samples.  For the early-type and intermediate-type galaxies, the
values of $\Phi_3(0.51)$ show a systematic difference of
$0.00006-0.0008$\phiunit~with the corresponding values of
$\Phi_1(0.51)$, due to the different weighting of the large-scale
structure by the 2 estimators (see \sct~\ref{phistar_equat}).  In
contrast, for the late-type galaxies, $\Phi_3(0.51)$ is systematically
lower than $\Phi_1(0.51)$ by $\sim0.001$\phiunit~ in $R_\mathrm{c}$ to
$\sim0.002$\phiunit~ in $V$ and $B$. Ignoring the evolution in the
late-type density and using the $\Phi_3$ estimator of $\phi_0$ rather
than $\Phi_1$ would then underestimate $\phi_0$ at $z\sim0.5$ by
$\sim40-50$\% in $R_\mathrm{c}$, and by $\sim70$\% in $V$ and $B$.

For the late-type galaxies, we also list in Table~\ref{phistar_BVR}
the linear parameterization of $\Phi_1(z)$ (\eqs~\ref{phi_evol_late}
and \ref{phi_015}). Comparison of the ``zero-point'' of the linear
parameterization $\Phi_1(0.15)$ with the listed value of
$\Phi_1(0.51)$ illustrates the increase in $\Phi_1$ from $z\sim0.15$
to $z\sim0.51$.  Note that Table~\ref{phistar_BVR} also lists the
parameters derived from the shallower but more complete
$R_\mathrm{c}\le20.5$ sample (using the same LFs as for the
$R_\mathrm{c}\le21.5$ sample, see Table~\ref{lf_BVR}). The evolution
rate $P_{0.15}=4.45$ is close to that for the deeper $R_\mathrm{c}$
sample, and the values of $\Phi_1$ and $\Phi_3$ differ from those from
the $R_\mathrm{c}\le21.5$ sample by less than 1-$\sigma$ (the
uncertainties in $\Phi_1$ and $\Phi_3$ for all considered samples in
Table~\ref{phistar_BVR} are in the interval $\sim0.0002-0.0003$ for
$z_\mathrm{max}\ge0.3$).

We apply a similar analysis to the ESS $V\le21.0$ and $B\le22.0$
samples. The measured and modeled relative $\Phi_1$ estimators of
$\phi_0$ are plotted in \fg~\ref{phistar_VB}; the relative $\Phi_1$
for the intermediate-type galaxies are also plotted in
\fg~\ref{phistar_VB}.  As in the $R_\mathrm{c}$ filter, the fits are
performed in the restricted redshift interval $0.15\le z\le0.5$.  In
both the $V$ and $B$ bands, we find an increase of the late-type
galaxy density which confirms the reality of the effect detected in
the $R_\mathrm{c}$ band.  We however measure a factor of 2 higher
evolution rate $P_{0.15}$ in $V$ and $B$ than in $R_\mathrm{c}$ (see
Table~\ref{phistar_BVR}). This may be due to the fact that when going
to bluer bands, the late-type galaxies are favored.  The higher
evolution rate in the $V$ and $B$ bands may also reflect the
color-dependent selection effects affecting the redshift samples in
the $V$ and $B$ bands. Moreover, the variations in $P_{0.15}$ from
band to band are symptomatic of the limited constraints provided by
the ESS \emph{redshift} distributions.  In the following, we show that
measuring the evolutions rates from the ESS faint magnitude number
counts yields better agreement among the 3 filters.

\section{Modeling the ESS number-counts \label{counts} }

\begin{figure}
  \resizebox{\hsize}{!}
    {\includegraphics{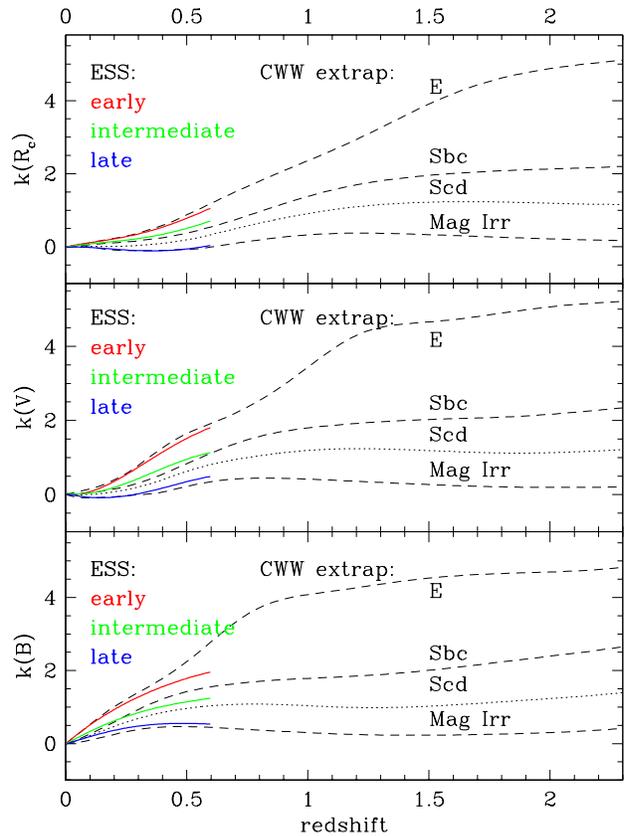}}
\caption{K-corrections in the $R_\mathrm{c}$ bands (top panel), $V$
(middle panel), and $B$ (bottom panel) derived from \citet[labeled
``CWW extrap'' in graph; see text for details]{coleman80} which are
used for predicting the ESO-Sculptor galaxy number counts in
\fgs~\ref{count_R}, \ref{count_V} and \ref{count_B}: Elliptical, Sbc
and Magellanic Irregular types are plotted as dashed lines (running
from top to bottom), and are used to model the ESO-Sculptor
K-corrections for the early-type, intermediate-type, and late-type
galaxies respectively; K-corrections for ``CWW extrap'' type Scd are
shown as dotted lines, for comparison.  The polynomial K-corrections
calculated for the 3 ESO-Sculptor spectral classes and valid only for
$z<0.6$ are shown as heavy solid lines, with early-type,
intermediate-type and late-type galaxies plotted from top to bottom.}
\label{kcor_z2}
\end{figure}

\begin{figure}
  \resizebox{\hsize}{!}
    {\includegraphics{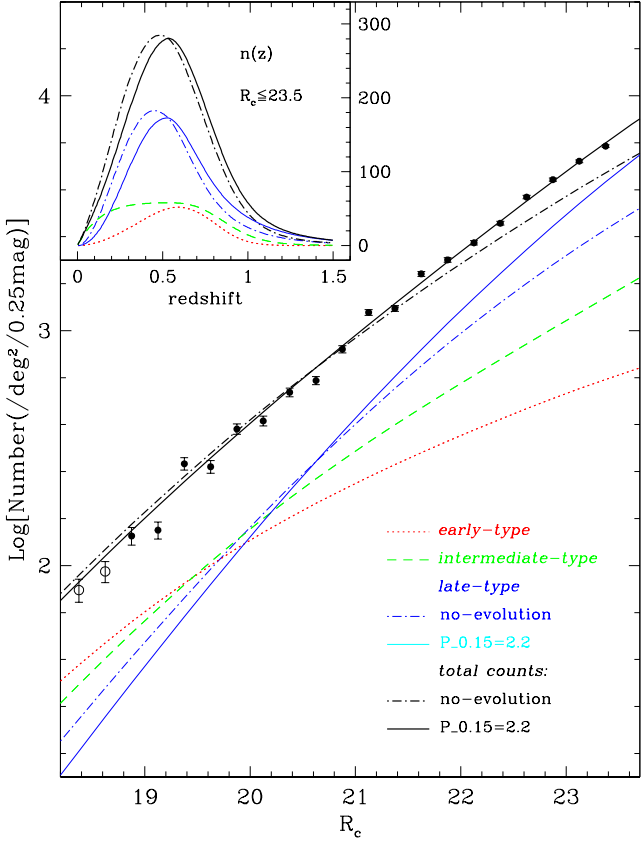}}
\caption{Comparison of the ESO-Sculptor magnitude number-counts in the
$R_\mathrm{c}$ band per 0.25\mag~interval (open and filled circles)
with the modeled counts defined as the sum of the predicted number
counts for the 3 considered spectral classes: early-type (red dotted
line), intermediate-type (green dashed line), late-type with
no-evolution (blue dot-dashed line), late-type with the best-fit
evolution rate (blue solid line). The Gaussian+Schechter composite
luminosity functions listed in Table~\ref{lf_BVR} are used for the 3
classes. The amplitude $\phi_0$ of the Gaussian component of the
luminosity function used for each galaxy class is identical to that
used in \fg~\ref{nz_R}.  The black heavy dot-dashed line show the
predicted total counts with no-evolution, and the black heavy solid
line the adjusted total counts with evolution in the late-type
galaxies (see text for details). Only the data points with more than
100 galaxies per bin (filled circles) are used for the adjustment of
the total evolving counts.  The plotted error bars in the observed
counts are estimated as $\sqrt{n}$. The inset shows the predicted
redshift distributions for the 3 spectral classes, with the same line
coding as for the number-counts.}
\label{count_R}
\end{figure}

\begin{figure}
  \resizebox{\hsize}{!}
    {\includegraphics{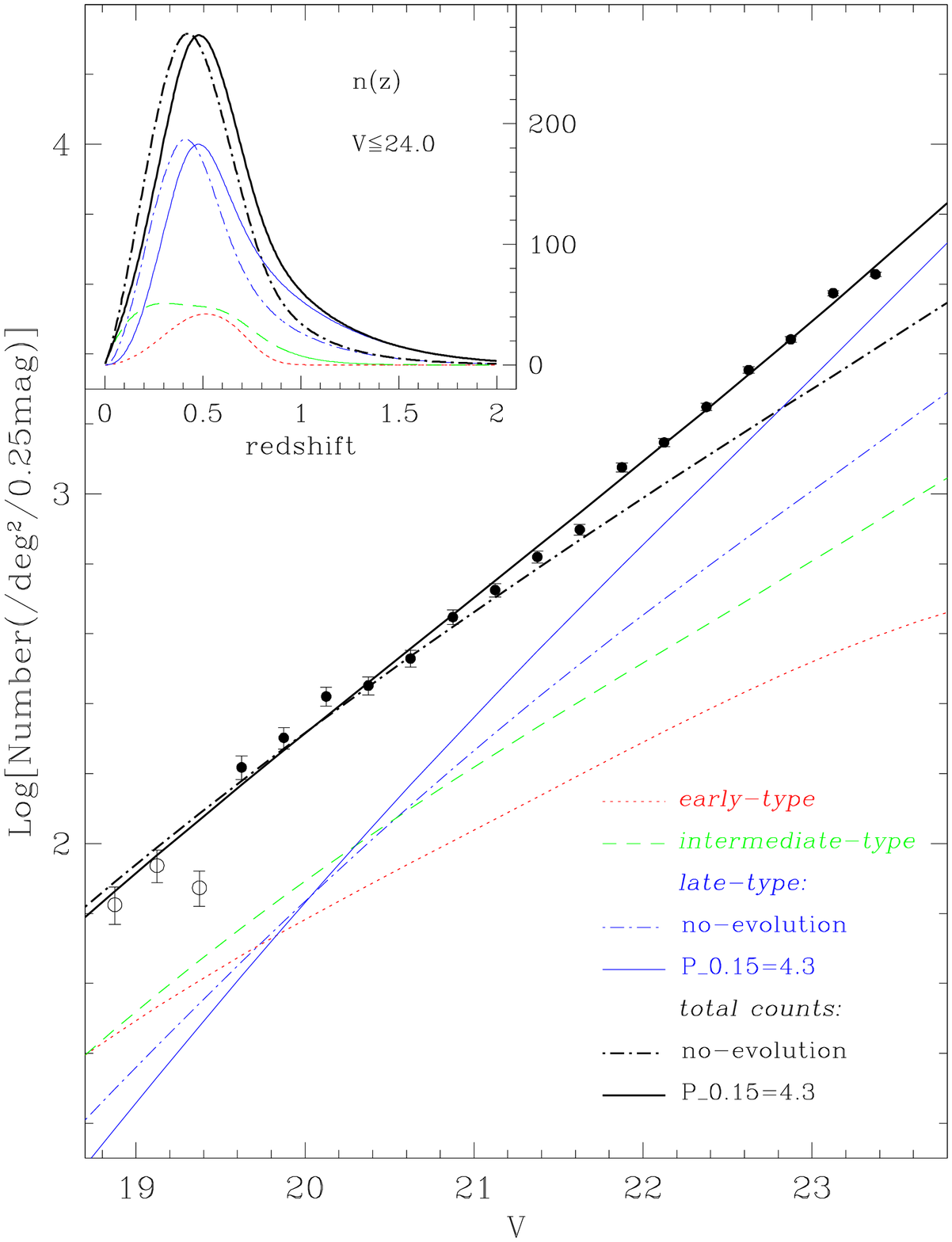}}
\caption{Same as \fg~\ref{count_R} for the ESS number-counts per 0.25
mag interval in the $V$ band. The amplitude $\phi_0$ of the Gaussian
component of the luminosity function used for each galaxy class is
identical to that used in \fg~\ref{nz_V}.}

\label{count_V}
\end{figure}

\begin{figure}
  \resizebox{\hsize}{!}
    {\includegraphics{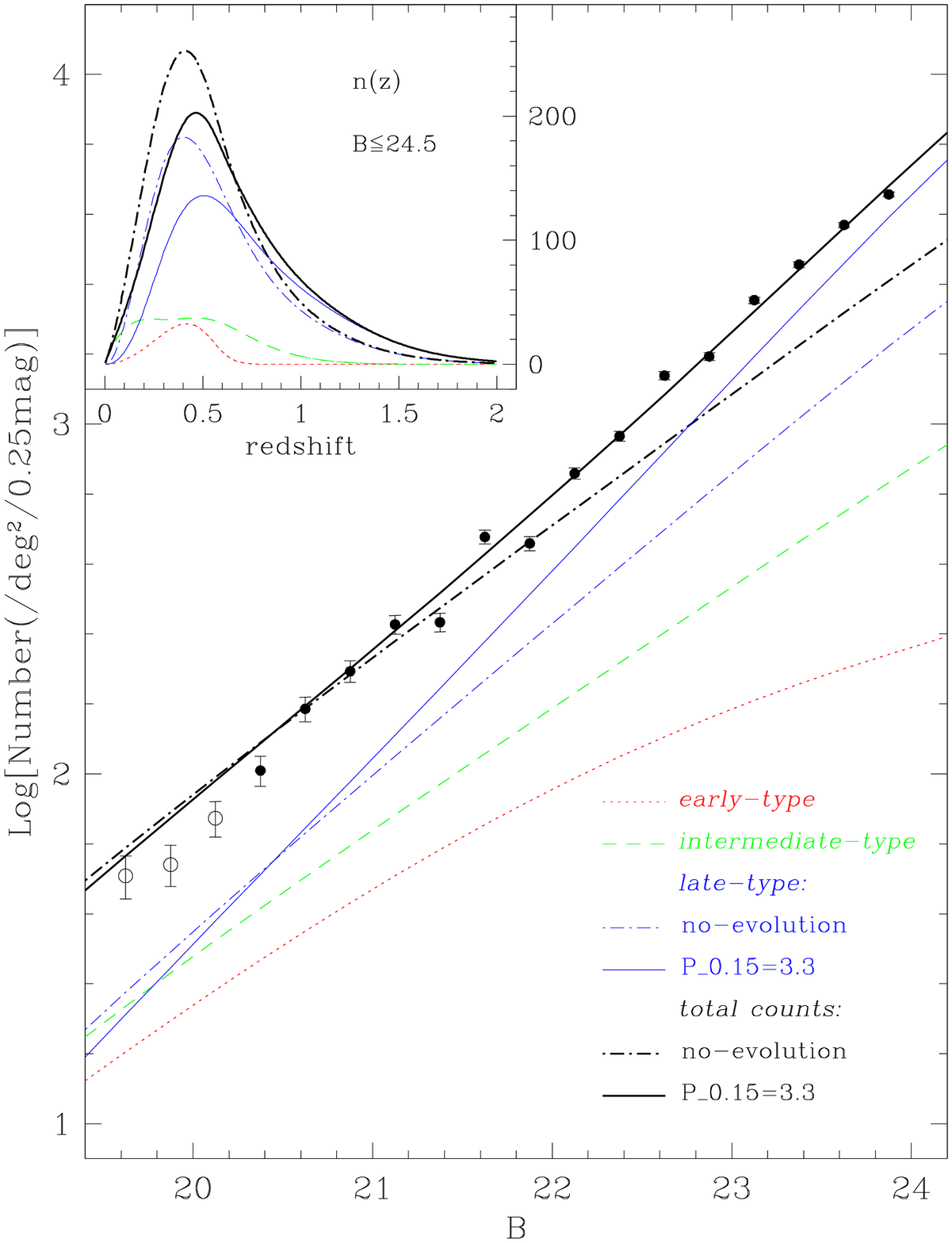}}
\caption{Same as \fg~\ref{count_R} for the ESS number-counts per 0.25
mag interval in the $B$ band. The amplitude $\phi_0$ of the Gaussian
component of the luminosity function used for each galaxy class is is
identical to that used in \fg~\ref{nz_B}.}
\label{count_B}
\end{figure}

Because the ESS number-counts extend nearly 3 magnitudes fainter than
the spectroscopic catalogue, comparison of the observed galaxy counts
with those predicted by the LFs provides a test of how well the
measured LFs and evolution rates can be extrapolated from $z\simeq0.5$
to $z\simeq1$.

Figures~\ref{count_R}, \ref{count_V}, \ref{count_B} show the observed
number counts in the $R_\mathrm{c}$, $V$ and $B$ bands \resp, binned
in intervals of 0.25\mag~\citep{arnouts97}. In each band, the
magnitude of the faintest plotted point corresponds to the
completeness limit: $R_\mathrm{c}\simeq23.5$, $V\simeq23.5$, and
$B\simeq24$, respectively. At bright magnitudes, the plotted number
count distributions start in the first 0.25\mag~bin where the count is
larger or equal to 50 galaxies, as the counts are highly uncertain at
low count level. For the same reason, the fits described in the
following start in the first bin where the count is larger or equal to
100, that is in the bins centered at $R_\mathrm{c}=18.875$, $V=19.625$
and $B=20.325$, respectively.

For modeling the galaxy number-counts, one must define a set of
K-corrections. We cannot use the ESS polynomials K-corrections
described in \sct~\ref{sp_kcor}, as these are unconstrained at
$z>0.6$, whereas the number-counts to magnitudes $\sim24$ are produced
by galaxies out to $z\sim 1.5$. In replacement, we use the
K-corrections obtained from the optical spectra of \citet{coleman80},
which have been extrapolated in the UV and the IR by using the
theoretical SEDs of the GISSEL library \citep{charlot96} for the 4
types Elliptical, Sbc, Scd, and Magellanic Irregular
\citep[see][]{sawicki97,arnouts99}. These K-corrections are shown in
\fg~\ref{kcor_z2} for the $B$, $V$ and $R_\mathrm{c}$ bands (labeled
``CWW extrap''). We choose to use for the ESS early-type,
intermediate-type, and late-type, the K-corrections for the
Elliptical, Sbc, and Magellanic Irregular types respectively. This
choice is motivated by the comparison for $z\le 0.6$ of the ``CWW
extrap'' K-corrections with the ESS K-corrections for the 3 spectral
classes, also shown in \fg~\ref{kcor_z2}. In the common redshift
interval, the 2 sets of K-corrections are in good agreement, with the
largest deviations occurring in the $B$ band ($\sim0.8$\mag~for the
early-type, $\simeq 0.3$\mag~for the intermediate type, and $\simeq
0.1$\mag~for the late-type galaxies), as it is the most sensitive band
to the template shape in the UV at the redshifts considered here.  The
consistent evolution rates obtained in the following from the
number-counts in the $R_\mathrm{c}$, $V$ and $B$ bands a posteriori
indicates that the CWW-extrap K-corrections do not introduce any
severe systematic effects in the modeled number-counts.

We then model the expected number-counts in each band as the sum
of the predicted number-counts over the 3 galaxy spectral classes
$T_\mathrm{S}$:
\begin{equation}
N(m)=\Sigma_{T_\mathrm{S}=1}^{3}N_{T_\mathrm{S}}(m),\\
\label{sum_count}
\end{equation}
where $m$ is the apparent magnitude in either of the $R_\mathrm{c}$,
$V$ or $B$ filters.  For each class, the expected number count
$N_{T_\mathrm{S}}(m)$ is obtained by integrating the corresponding
composite LF $\Phi_{T_\mathrm{S}}(M)$ (listed in Table~\ref{lf_BVR})
over all redshifts $0\le z\le z_\mathrm{high}$ contributing to
absolute magnitude $M(m,z,T_\mathrm{S})$ at a fixed value of $m$ (see
\eq~\ref{abs_mag}):
\begin{equation}
N_{T_\mathrm{S}}(m)=\int_0^{z_\mathrm{high}}
\Phi_{T_\mathrm{S}}[M(m,z,T_\mathrm{S})]\;
\frac{\mathrm{d}V}{\mathrm{d}z}\;\mathrm{d}z;\\
\label{type_count}
\end{equation}
$\frac{\mathrm{d}V}{\mathrm{d}z}$ is the comoving volume element per
unit solid angle defined as
\begin{equation}
\frac{\mathrm{d}V}{\mathrm{d}z}=
\frac{c}{H_0}\;\frac{d_\mathrm{L}(z)^2}{(1+z)^2}\;
\frac{1}{\sqrt{\Omega_\mathrm{m}(1+z)^3+\Omega_\Lambda}}.
\label{dv}
\end{equation}
The values of $z_\mathrm{high}$ are $1.5$ in the $R_\mathrm{c}$ band,
and $2.0$ in the $B$ and $V$ bands.  Note that $N(m)$ and
$N_{T_\mathrm{S}}(m)$ in \eqs~\ref{sum_count} and \ref{type_count}
\resp~ are defined as number counts per unit solid angle and per unit
magnitude interval. The curves plotted in \fgs~\ref{nz_R} to
\ref{nz_B} are then multiplied by 0.25\mag~and $\pi^2/180^2$ (to
convert to number counts per 0.25\mag~interval per square degree).

\begin{table*}
\caption{Normalizing amplitudes $\phi_0$ and $\phi^*$ of the Gaussian
and Schechter components of the early-type, intermediate-type and
\emph{non-evolving} late-type luminosity functions of the ESO-Sculptor
survey, in the Johnson-Cousins $B$, $V$ and $R_\mathrm{c}$ bands.}
\label{phistar_noevol}
\begin{center}
\begin{tabular}{lcccccccccc}
\hline 
\hline 
& \multicolumn{5}{l}{($\Omega_m$,$\Omega_\Lambda$)=(0.3,0.7)} & \multicolumn{5}{l}{($\Omega_m$,$\Omega_\Lambda$)=(1.0,0.0)}\\ 
              & Early-type & \multicolumn{2}{c}{Intermediate-type} & \multicolumn{2}{c}{Late-type} & Early-type & \multicolumn{2}{c}{Intermediate-type}& \multicolumn{2}{c}{Late-type}\\
                     & $\phi_0$  & $\phi_0$  & $\phi^*$  & $\phi_0$  & $\phi^*$ & $\phi_0$  & $\phi_0$  & $\phi^*$  & $\phi_0$  & $\phi^*$  \\
\hline
$R\mathrm{c}\le21.5$ & $0.00310$ & $0.00296$ & $0.00387$ & $0.00339$ & $0.03678$ & $0.00504$ & $0.00489$ & $0.00639$ & $0.00520$ & $0.05649$ \\
$V\le21.0$           & $0.00323$ & $0.00287$ & $0.00375$ & $0.00271$ & $0.02939$ & $0.00497$ & $0.00444$ & $0.00581$ & $0.00410$ & $0.04450$ \\
$B\le22.0$           & $0.00344$ & $0.00294$ & $0.00385$ & $0.00299$ & $0.03247$ & $0.00523$ & $0.00458$ & $0.00600$ & $0.00457$ & $0.04961$  \\
\hline
\end{tabular}
\smallskip
\end{center}

\begin{list}{}{}
\item[\underline{Notes:}]
\item[-]The listed values of $\phi_0$ and $\phi^*$ are in units of
\phiunit, and are obtained with $H_0 = 100 h$ km s$^{-1}$ Mpc$^{-1}$.
\item[-]$\phi_0$ and $\phi^*$ are calculated
by normalizing the integral of the expected redshift distribution with
the listed apparent magnitude limit to the observed number of galaxies
in the interval $0.01\le z\le0.81$ (see \sct~\ref{nz}).
\item[-]The value of $\phi^*$ for the \emph{intermediate-type}
and \emph{late-type} galaxies is related to the corresponding value of
$\phi_0$ by the ratio $\phi_0/0.4\ln10\;\phi^*$, listed in
Table~\ref{lf_BVR}.
\end{list}
\end{table*}

\begin{table*}
\caption{Evolution rates and zero-points for the amplitudes $\phi_0$
and $\phi^*$ of the Gaussian and Schechter components of the late-type
luminosity function in the ESO-Sculptor Survey, as derived from the
number-counts in the Johnson $B$, $V$ and Cousins $R_\mathrm{c}$
bands.}
\label{evol_BVR}
\begin{center}
\begin{tabular}{lcccccccccccc}
\hline 
\hline 
& \multicolumn{6}{l}{($\Omega_m$,$\Omega_\Lambda$)=(0.3,0.7)} & \multicolumn{6}{l}{($\Omega_m$,$\Omega_\Lambda$)=(1.0,0.0)}\\ 
            &  $P_{0.15}$ & $\phi_0(0.15)$ & $\phi^*(0.15)$ & $\gamma$ & $\phi_0(0)$ & $\phi^*(0)$ & $P_{0.15}$ & $\phi_0(0.15)$ & $\phi^*(0.15)$ & $\gamma$ & $\phi_0(0)$ & $\phi^*(0)$  \\
\hline
$R_\mathrm{c}$ & $2.2$ & $0.00252$ & $0.02732$ & $1.8$ & $0.00211$ & $0.02292$ & $4.6$ & $0.00320$ & $0.03470$ & $2.5$ & $0.00287$ & $0.03119$ \\
$V$            & $4.3$ & $0.00216$ & $0.02347$ & $2.5$ & $0.00186$ & $0.02019$ & $5.7$ & $0.00322$ & $0.03493$ & $2.8$ & $0.00283$ & $0.03072$ \\
$B$            & $3.3$ & $0.00267$ & $0.02899$ & $2.2$ & $0.00226$ & $0.02449$ & $5.0$ & $0.00355$ & $0.03857$ & $2.6$ & $0.00313$ & $0.03399$ \\
\hline
\end{tabular}
\smallskip
\end{center}

\begin{list}{}{}
\item[\underline{Note:}]
\item[-]The listed values of $\phi_0$ and $\phi^*$ are in units of
\phiunit, and are obtained with $H_0 = 100 h$ km s$^{-1}$ Mpc$^{-1}$.
\item[-]The linear, power-law parameterization of the evolution in $\phi_0$ and $\phi^*$
is defined in \eqs~\ref{phi_evol_late}, \ref{phi_evol_power}
respectively.
\item[-]Each value of $\phi^*$ is related to the corresponding
value of $\phi_0$ by the ratio $\phi_0/0.4\ln10\;\phi^*=0.1$ (see
Table~\ref{lf_BVR}).
\end{list}
\end{table*}

In \fgs~\ref{count_R}, \ref{count_V}, and \ref{count_B}, the thin
dotted, dashed, and dot-dashed lines correspond to the predicted
counts $N_{T_\mathrm{S}}(m)$ for the early-type, intermediate-type,
and late-type galaxies without evolution, and using
($\Omega_m$,$\Omega_\Lambda$)=(0.3,0.7). The amplitudes $\phi_0$ and
$\phi^*$ of the composite LFs are those which match the integrals of
the observed and predicted redshift distributions in the interval
$0.01\le z\le0.81$; these values are listed in
Table~\ref{phistar_noevol}, and are also indicated in \fgs~\ref{nz_R},
\ref{nz_V} and \ref{nz_B} (see \sct~\ref{nz}). The summed
number-counts over the 3 classes are plotted as heavy dot-dashed lines
in \fgs~\ref{count_R}, \ref{count_V}, and \ref{count_B}.  The
total non-evolving counts only match the observed number-counts at
magnitudes brighter than $R_\mathrm{c}\simeq21.5$, $V\simeq21.5$ and
$B\simeq22.0$ \resp, which corresponds or is close to the magnitude
limit of the respective redshift samples. At the faint limit, the
no-evolution counts under-predict the observed counts by $\sim 25$\%
in the $R_\mathrm{c}$ band, and by $\sim 60$\% in the $V$ and $B$
bands.

Moreover, there is no value of non-evolving amplitude for the
late-type LF which can both match the bright and faint ESS
number-counts in all 3 bands.  A simple scaling of the amplitude of
the late-type LFs by a factor $1.37$, $1.85$ and $1.90$ in the
$R_\mathrm{c}$, $V$ and $B$ bands \resp, obtained by a weighted
least-square minimization of the summed counts to the observed counts,
still fails to match simultaneously the observed number-counts at
bright and faint magnitudes in all 3 bands: this scaling makes little
change to the slope of the total counts, and essentially shifts them
upward.

Figures~\ref{count_R}, \ref{count_V} and \ref{count_B} show that in each
band, the number counts at magnitudes fainter than the limit of the
redshift sample ($\sim21-22$\mag) are dominated by the late-type
galaxies. Introducing a scaling factor or some evolution in the
amplitude of the early-type and/or intermediate-type LFs would
therefore bring no improvement in matching the observed faint
number-counts.  Better adjustments may only be obtained by increasing
the contribution from the late-type galaxies at faint magnitude. This
provides further evidence that the late-type galaxies are evolving.
In the following, we show that by introducing a linear (or power-law)
evolution in the amplitude of the late-type LF, one obtains a very
good adjustment of the observed number counts in the 3 filters. Note
that we deliberately do not consider any evolution in the early-type
and intermediate-type ESS galaxy populations, for the following
reasons:
\begin{itemize}
\item there are several indications of a marked decrease in the number
density of E galaxies with redshift \citep{fried01,wolf03} which
compensates for their passive luminosity evolution and explains that
no evolution is detected by the ESS in this population.
\item as far as the ESS intermediate spectral class is considered,
although a significant brightening due to passive evolution is
expected for galaxies with present-day colors resembling those of Sb
to Sbc morphological types (using intermediate values between those
provided for Sa and Sc galaxies by \citealt{poggianti97}), little or
no evolution in their luminosity density is detected out to
$z\sim0.5-1.0$ \citep{lin99,wolf03}.
\end{itemize}

We now determine the optimal evolution rate $P_{0.15}$ defined in
\eq~\ref{phi_evol_late} for the late-type galaxies by a 2-stage
procedure.  We first vary the value of $P_{0.15}$ and perform a
weighted least-square fit of the summed predicted counts to the
observed counts, with the amplitude of the early-type and
intermediate-type LFs kept fixed.  The weights are defined as the
square-root of the observed total counts, and might thus
underestimates the true uncertainty in the observed counts, which
should also account for galaxy clustering; we however verified that
increasing the \rms errors by as much as a factor of 2, a wide
overestimate of cosmic variance over the area of the ESS, would make
negligeable change in the derived evolution rates.  For each value of
$P_{0.15}$, the amplitude of the late-type LF is defined by matching
the observed and predicted late-type redshift distribution in the
interval $0.01\le z\le0.81$ (see next \sct).  The reference amplitude
of the late-type LF is therefore a function of $P_{0.15}$.  As this
value may however not provide the optimal match between the predicted
and observed late-type counts, we allow in each least-square fit for a
scaling factor to the amplitude of the late-type LF.  The full
procedure yields a first estimate of the evolution parameter
$P_{0.15}$ (defined as the value for which the reduced $\chi^2$ is
smallest): $2.2$, $4.3$ and $3.3$ in the $R_\mathrm{c}$, $V$, and $B$
bands \resp, with scaling factors $0.931$, $1.005$ and $1.1287$ for
the late-type LF amplitude.  The predicted number-counts for the
late-type galaxies with the above quoted evolution rates and scaling
factors are plotted as a light solid line in \fgs~\ref{count_R},
\ref{count_V} and \ref{count_B}, and the corresponding summed counts
over the 3 galaxy types as a heavy solid line.

We have also looked for other minima of the reduced $\chi^2$ by
allowing for a scaling factor in the amplitude of the early-type and
intermediate-type LFs, and searching for a \emph{common} scaling
factor for the 3 galaxy types. For values of $P_{0.15}$ around the
first minima listed above, we iterate over the values of the scaling
factors for the 3 classes: the output scaling factors for the
late-type galaxies are applied to both the early-type and
intermediate-type LF amplitudes, and the new scaling factor for the
late-type galaxies which minimizes the $\chi^2$ is calculated. After 5
to 10 iterations, this converges to a common scaling factor for the 3
galaxy types. When considering the final reduced $\chi^2$ obtained by
these iterations, a slightly smaller evolution rate $P_{0.15}=2.0$ is
obtained for the $R_\mathrm{c}$ counts, with a common scaling factor
of $0.972$; the same minimum $P_{0.15}=4.3$ is confirmed in the $V$
band, with a common scaling factor of $1.003$ for the 3 galaxy types;
and a slightly higher evolution rate $P_{0.15}=3.8$ is obtained in the
$B$ band, with a common scaling factor $1.055$. Note that in each
filter, the common scaling factor for the 3 galaxy types is closer to
unity than the factor used when scaling only the late-type LF. The
predicted number-counts with a common scaling factor are
indistinguishable from those in which only the late-type LFs are
scaled (which are shown in \fgs~\ref{count_R}-\ref{count_B}).

The excellent adjustment of the number counts using the linear
evolution model of the amplitude of the late-type LF while keeping
nearly constant the density of early-type and intermediate-type
galaxies provides evidence that the late-type galaxies evolve out to
$z\sim1.0$. The inserts of \fgs~\ref{count_R}, \ref{count_V} and
\ref{count_B} show that in the 3 bands, the expected redshift
distribution of the galaxies detected in the ESS number counts have a
peak near $z=0.5$ and extend to $z\ga1$.  Note also that the number
counts provide better agreement among the evolution rates in the 3
bands ($P_{0.15}=2.0-2.2$ in $R_\mathrm{c}$, $P_{0.15}=4.3$ in $V$,
$P_{0.15}=3.3-3.8$ in $B$) that those measured from the redshift
survey only ($P_{0.15}=3.5$ in $R_\mathrm{c}$, $P_{0.15}=7.5$ in $V$,
$P_{0.15}=8.6$ in $B$; see \sct~\ref{evol}).

We estimate the uncertainty in the value of $P_{0.15}$ measured from
the number-counts by applying the following tests: (i) in the
late-type LF, we change alternatively the slope $\alpha$ of the
Schechter component from $-0.3$ to $0.39$, the value actually measured
from the $R_\mathrm{c}\le21.5$ sample (see Paper~I), and the peak
magnitude $M_0$ of the Gaussian component by $\pm0.3$\mag, as these
are the 2 parameters which have the largest impact on $P_{0.15}$; (ii)
we vary the amplitude of the early-type or intermediate-type LFs by
$\pm10$\%, which provides a conservative estimate of the uncertainties
in the values of $\phi^*$ and $\phi_0$ for these samples (see
\eqs~\ref{sigma_n3}-\ref{sigma_n2}). Each of these tests yields a
change in $P_{0.15}$ by $\la0.5$. We thus adopt as a conservative
uncertainty $\sigma(P_{0.15})\simeq1.0$.  The above values of
$P_{0.15}$ obtained from the number-counts then differ by $1-2\sigma$
from filter to filter.

We emphasize that despite the incompleteness in blue galaxies in the
ESS $V$ and $B$ \emph{spectroscopic} samples, when used together with
the ESS $V$ and $B$ magnitude \emph{number-counts}, they do provide
useful constraints on the evolution rate for the late-type galaxies,
and yield consistent results with those derived from the
$R_\mathrm{c}$ number-counts. Note that in the estimation of
$P_{0.15}$ from the number-counts, the $V$ and $B$ spectroscopic
samples are used only to derive the amplitude of the LFs by
normalizing to the observed redshift distributions.  The consistent
evolution rates obtained in the $B$, $V$, and $R_\mathrm{c}$ bands
reinforces the detected evolution as a real effect.  The tendency of
an increased evolution rate measured in the $V$ and $B$ bands compared
to the $R_\mathrm{c}$, may be due to the higher sensitivity of the $V$
and $B$ bands to the late-type galaxies, already mentioned in
\sct~\ref{evol}: at the peak redshift probed by the ESS number-counts
($z\sim0.5$), taken together the $B$ and $V$ bands probe the
rest-wavelength interval $\sim2700-4000$\AA, lying just blue-ward of
the CaII H \& K break; the significant star formation activity present
in late-type galaxies does cause an increased flux at these
wavelengths.

Note that applying the above analysis using pure Schechter LFs with a
common slope $\alpha=-1.48$ for the $R_\mathrm{c}$, $V$, and $B$ LFs
(Paper~I) yields values of $P_{0.15}$ smaller by 0.5 than those
derived from the composite LFs. However, the use of pure Schechter LFs
yields a marked degeneracy between the slope $\alpha$ of the late-type
LF and the evolution rate $P_{0.15}$.  For example, changing the slope
$\alpha$ of the late-type LF from $-1.64$, measured from the
$R_\mathrm{c}\le20.5$ sample, to $-1.48$, measured from the
$R_\mathrm{c}\le21.5$ sample (Paper~I) yields an increase in
$P_{0.15}$ by nearly one unit. This is to be contrasted with the
change of only 0.3 in $P_{0.15}$ obtained when changing the slope of
the Schechter component of the late-type composite LF from $-0.3$ to
$0.39$ (note the large variation). The degeneracy in the faint-end
slope $\alpha$ of a Schechter LF could be partially reduced using the
redshift distribution, but there remains large uncertainties, as the
ESS spectroscopic sample is far from a fair sample of Universe, and
the redshift distribution fails in averaging out the large-scale
structure (see \sct~\ref{nz}).

In Table~\ref{evol_BVR}, we list the values of $P_{0.15}$ obtained in
the 3 filters for ($\Omega_m$,$\Omega_\Lambda$)=(0.3,0.7); there, the
secondary iteration stage aimed at obtaining a common scaling factor
for the 3 spectral classes is not used, as it makes a small difference
in the evolution parameter. We also measure $P_{0.15}$ for
($\Omega_m$,$\Omega_\Lambda$)=(1.0,0.0); in that case, the LF
characteristic magnitudes $M_0$ and $M^*$ listed in Table~\ref{lf_BVR}
are shifted by $\Delta M=0.3 ^\mathrm{mag}$, the variation in absolute
magnitude corresponding to the change in luminosity distance at
$z=0.3$ (the approximate peak redshift of the ESS; see
\sct~\ref{shape}). The corresponding values of $\phi_0$ and $\phi^*$
for the early-type, intermediate-type and non-evolving late-type LFs,
obtained by normalizing the integral of the expected redshift
distribution with ($\Omega_m$,$\Omega_\Lambda$)=(1.0,0.0) to the
observed number of galaxies in the interval $0.01\le z\le0.81$ are
also listed in Table~\ref{phistar_noevol}.

Finally, we apply the power-law evolution model of
\eq~\ref{phi_evol_power} to the number-counts and derive the best-fit
value of $\gamma$ for both sets of cosmological parameters; the
resulting values of $\gamma$ are listed in Table~\ref {evol_BVR}.  As
for the $P_{0.15}$ parameter, the uncertainty in $\gamma$ is estimated
to be of order of $1.0$. In the 3 bands, the minimum $\chi^2$ is
systematically smaller for the linear evolution model than for the
power-law model, but the difference is small. Note that the larger
evolution parameters obtained with cosmological parameters
($\Omega_m$,$\Omega_\Lambda$)=(1.0,0.0) are due to the corresponding
smaller volume element at increasing $z$.

\section{The ESS redshift distributions per spectral-type \label{nz} }

\begin{figure}
  \resizebox{\hsize}{!}
    {\includegraphics{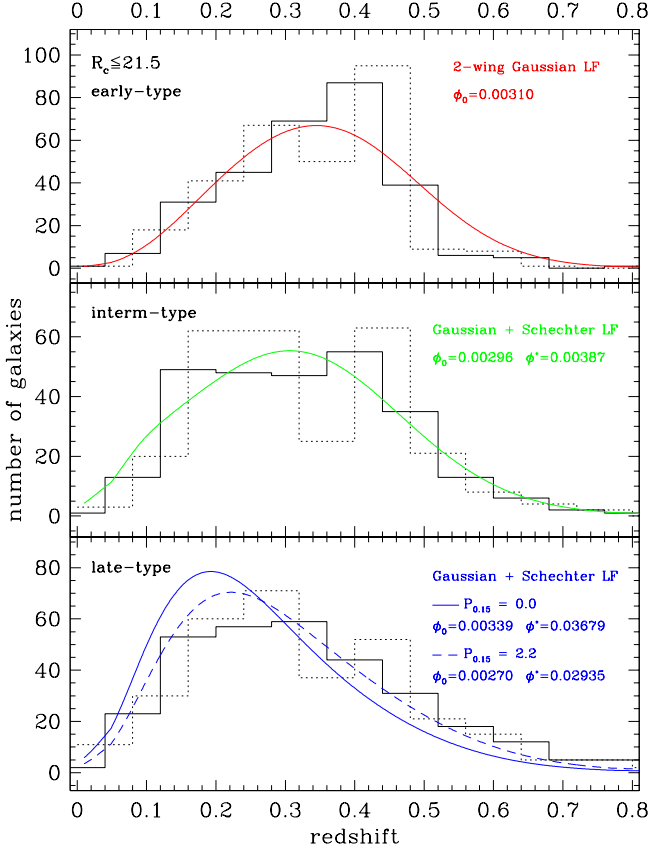}}
\caption{The observed redshift distribution for the early-type (top
graph), intermediate-type (middle graph), and late-type galaxies
(bottom graph) in the ESS $R_\mathrm{c}\le21.5$ sample. In order to
allow for the noise in the observed histograms, these are plotted for
2 binnings of $\Delta z=0.08$, offset by $0.04$ in redshift (first bin
starting at $z=0$, $z=0.04$ for the dotted, resp.  solid
histogram). The expected distribution for a uniform galaxy
distribution with the ESS incompleteness and the composite luminosity
function listed in Table~\ref{lf_BVR} is over-plotted as a solid line
for the 3 spectral classes. For the late-type galaxies, we also plot
as a dashed line the expected distribution with an evolution factor
$P_{0.15}=2.2$, as derived from the $R_\mathrm{c}$ number counts (see
\sct~\ref{counts}). All expected curves are normalized so that they
match the observed number of galaxies in the interval $0.01\le
z\le0.81$; the resulting amplitudes $\phi_0$ and $\phi^*$ for the
Gaussian and Schechter components are indicated inside each panel.}
\label{nz_R}
\end{figure}

\begin{figure}
  \resizebox{\hsize}{!}
    {\includegraphics{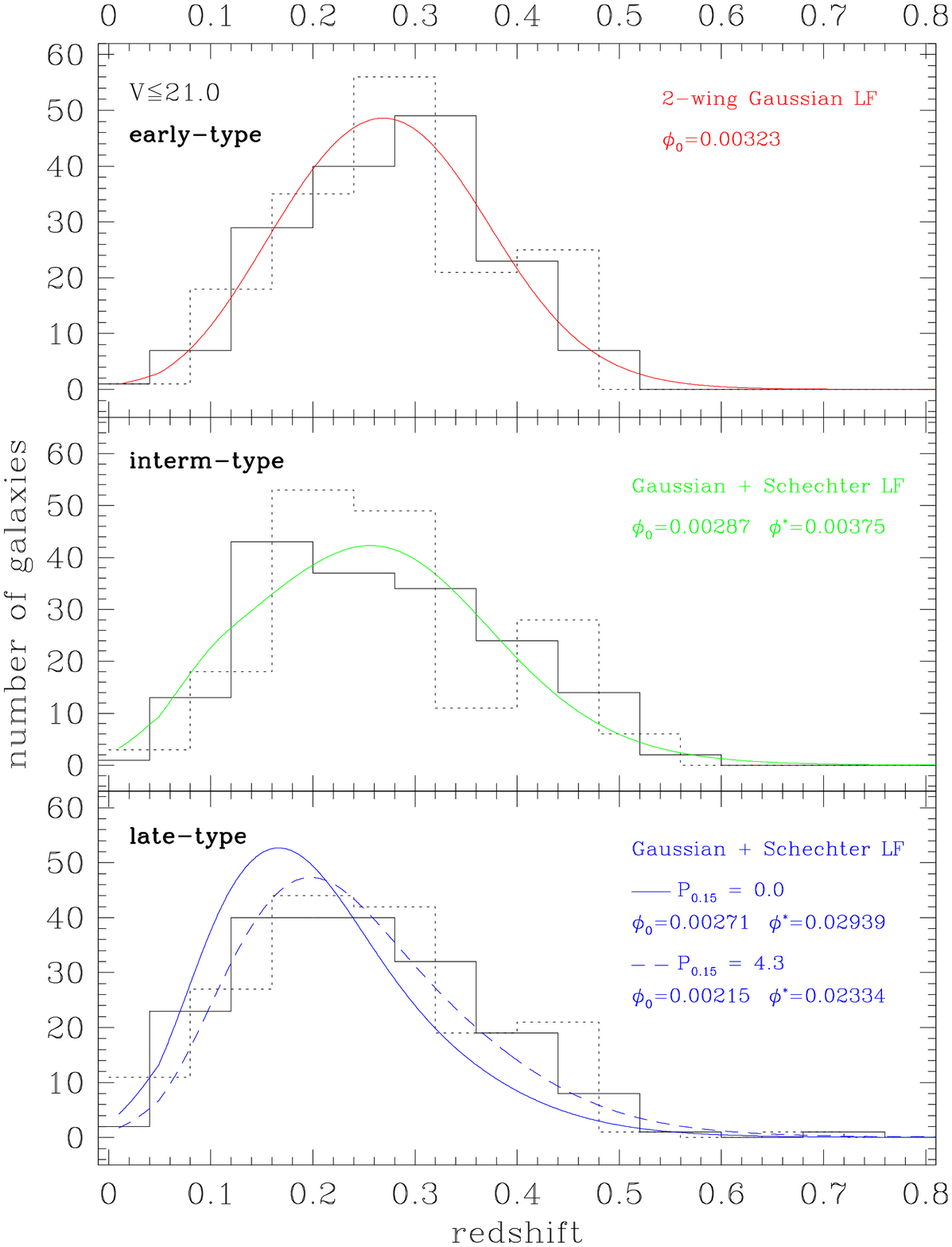}}
\caption{Same as \fg~\ref{nz_R} for the ESS $V\le21.0$ sample, with
$P_{0.15}=4.3$ as derived from the $V$ number counts (see
\sct~\ref{counts}).}
\label{nz_V}
\end{figure}

\begin{figure}
  \resizebox{\hsize}{!}
  {\includegraphics{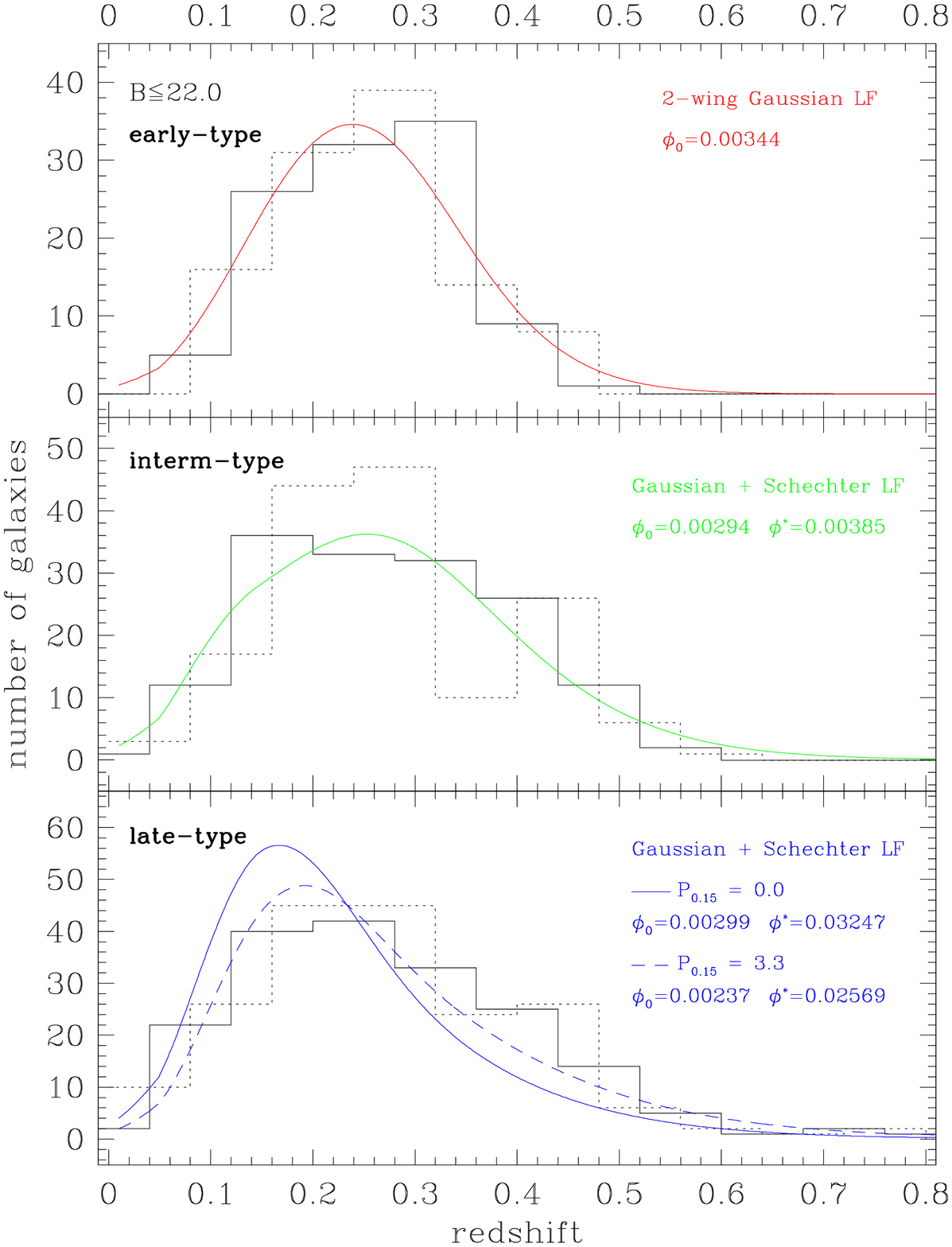}}
\caption{Same as \fg~\ref{nz_R} for the ESS $B\le22.0$ sample, with
$P_{0.15}=3.3$ as derived from the $B$ number counts (see
\sct~\ref{counts}).}
\label{nz_B}
\end{figure}

In \fgs~\ref{nz_R}, \ref{nz_V}, and \ref{nz_B}, we compare the
observed redshift distributions for the 3 spectral classes in the
$R_\mathrm{c}\le21.5$, $V\le21.0$ and $B\le22.0$ samples \resp~with
the expected distributions calculated using the composite LFs listed
in Table~\ref{lf_BVR}.  For the observed distributions, we plot the 2
histograms obtained with a redshift bin $\Delta z=0.08$ and offset by
$0.04$ in redshift, in order to illustrate visually the uncertainties
in the observed distribution.  A large bin size in redshift is used in
order to smooth out the variations due to large-scale clustering; this
scale would correspond to $\sim 120$\hmpc~ at small redshift, larger
than the typical size of the voids in the redshift surveys to
$z\le0.1$ \citep{lapparent86,shectman96,small97a,colless01,zehavi02},
and comparable to the scale of the largest inhomogeneities detected so
far in redshift surveys \citep{broadhurst90,geller97}.  The marked
deviations between the 2 histograms are due to large-scale structure
on even larger scales: a deficit of observed galaxies in the interval
$0.33\la z\la0.39$, and an excess in the interval $0.39\la z\la0.46$;
it is however unclear whether these structures extend beyond the
limited angular scale of the ESS.

As in \sct~\ref{phistar_equat}, the expected curves in
\fgs~\ref{nz_R}, \ref{nz_V}, and \ref{nz_B} are based on the integral
of the selection function over the bin-size $\Delta z=0.08$, and we
use the K-corrections calculated for the average spectral-type
$<T_\mathrm{S}>$ among each spectral class (see \eq~\ref{abs_mag}).
For the 3 galaxy types and in the 3 filters, the amplitudes $\phi_0$
and $\phi^*$ defining each non-evolving expected curve are defined by
normalizing the integral of the expected distribution to the observed
number of galaxies in the interval $0.01\le z\le0.81$ (the ratio
$\phi_0/0.4 \ln10\;\phi^*$ takes the values listed in
Table~\ref{lf_BVR}). The resulting values of $\phi_0$ and $\phi^*$ are
indicated inside each graph of \fgs~\ref{nz_R} to \ref{nz_B}, and
are also listed in Table~\ref{phistar_noevol} in column labeled
($\Omega_m$,$\Omega_\Lambda$)=(0.3,0.7). These values of $\phi_0$
differ from the estimates $\Phi_3(0.51)$ listed in
Table~\ref{phistar_BVR}, because the latter result from an integral
over the narrower redshift interval $0.1\le z\le0.51$ (see
\sct~\ref{evol}). 

For the early-type and intermediate-type galaxies, the expected
distributions in \fgs~\ref{nz_R}, \ref{nz_V} and \ref{nz_B} provide a
good match to the observed histograms. For the late-type galaxies, the
expected no-evolution redshift curves (with $P_{0.15}=0.0$) show a
systematic shift towards low redshifts when compared to the observed
distributions.  Moreover, the expected curves lie systematically near
the lower values of the observed histograms for $z\ga0.3$.  These
effects are present in the 3 filters.

For the late-type galaxies, we also plot in \fgs~\ref{nz_R},
\ref{nz_V} and \ref{nz_B} the expected redshift distributions with the
values of the evolution factor $P_{0.15}$ listed in
Table~\ref{evol_BVR} for ($\Omega_m$,$\Omega_\Lambda$)=(0.3,0.7).  The
values of $\phi_0(0.15)$ and $\phi^*(0.15)$ which normalize the
integral of each evolving distribution to the corresponding observed
number of galaxies in the interval $0.01\le z\le0.81$ are indicated in
\fgs~\ref{nz_R} to \ref{nz_B}. Note that these values differ from
those listed in Table~\ref{evol_BVR}, as the latter are derived by
normalization to the total number-counts. The difference is however
small, $\la10$\%, thus bringing a posteriori evidence of consistency
between the redshift and magnitude distributions.  Note also that in
the evolving curves, we have extrapolated to $0\le z\le0.15$ the
linear evolution of $\Phi$ for the late-type galaxies parameterized in
\eq~\ref{phi_evol_late}, although it was measured from the restricted
redshift interval $0.15 \le z \le 0.5$. Our motivations for this
choice are:
\begin{itemize}
\item if we assume that $\Phi$ remains constant and at the value of
the linear fit at $z=0.15$ for redshifts $\le0.15$, the expected
redshift distributions show an marked excess of galaxies at these
redshifts, which does not match the observed distribution (this effect
is observed in all filters).
\item the location of the ESS was visually selected by examining
copies of the ESO/SERC R and J atlas sky survey with the criteria to
avoid nearby galaxies and nearby groups and clusters of galaxies. This
implies that the ESS survey has a systematically low density of
galaxies at $z\la0.1$.
\end{itemize}

\begin{figure}
  \resizebox{\hsize}{!}
    {\includegraphics{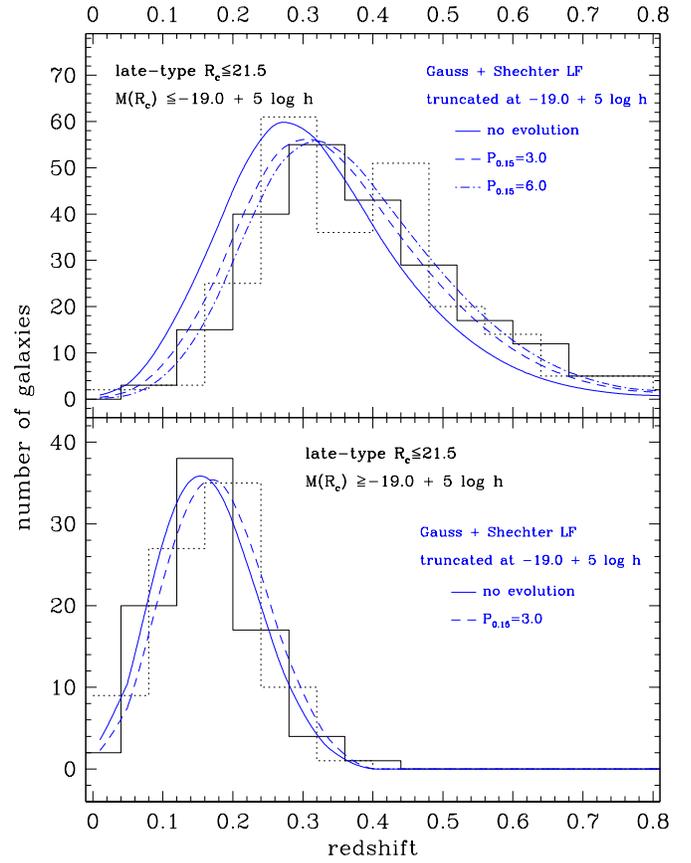}}
\caption{The observed redshift distribution for the bright and faint
galaxies in the ESO-Sculptor late-type class
($M(R_\mathrm{c})\le-19.0+5\log h$ in top panel,
$M(R_\mathrm{c})\ge-19.0+5\log h$ in bottom panel).  In order to allow
for the noise in the observed histograms, they are plotted for 2
binnings of $\Delta z=0.08$, offset by $0.04$ in redshift (first bin
starting at $z=0$, $z=0.04$ for the dotted, \resp~solid
histogram). The expected distribution for a uniform galaxy
distribution with the ESS incompleteness and the composite late-type
LF listed in Table~\ref{lf_BVR} and truncated at
$M(R_\mathrm{c})=-19.0+5\log h$ is plotted as a solid line in both panels.
The expected distribution with an evolution factor $P_{0.15}=3.0$ is
plotted as a dashed line in both panels; the expected curve with
$P_{0.15}=6.0$ is also shown as a dash-dotted line in the top
panel. The expected curves are normalized so that they match the
observed number of galaxies in the interval $0.01\le z\le0.81$.}
\label{nz_ScIm}
\end{figure}

In contrast to the no-evolution curves, the evolving late-type
distributions provide a good match to the observed histograms in
\fgs~\ref{nz_R}, \ref{nz_V} and \ref{nz_B}: the low redshift peak is
shifted to higher redshift ($z\ga0.2$), and the high redshift tail
has a higher amplitude which better matches the observed data. The
best agreement of the observed and evolving expected curve is obtained
in the $R_\mathrm{c}$ band (\fg~\ref{nz_R}). In the $V$ and $B$ bands
(\fgs~\ref{nz_V} and \ref{nz_B}), the expected curves with
$P_{0.15}=4.3$ and $P_{0.15}=3.3$ \resp~may be still be too low at
$z\ga0.3$.  Using $P_{0.15}=7.0$ and $P_{0.15}=8.0$ in the $V$ and $B$
bands (as obtained from the direct fits of $\Phi_1$ in
\sct~\ref{evol}) yields a better match of the observed redshift
distributions at $z\ga0.3$, but systematically underestimate the
distributions at $z\la0.15$. This may be an indication that
extrapolation of the linear model of \eq~\ref{phi_evol_late} to $z\le
0.15$ is not satisfying for large values of $P_{0.15}$. The varying
incompleteness with galaxy type which affects the $V$ and $B$ samples
is also likely to complicate the adjustment of the redshift
distributions.

Note that the incompleteness of the \emph{full} sample is used to
correct for the incompleteness in apparent magnitude of each
\emph{spectral-type} sample in \fgs~\ref{nz_R}, \ref{nz_V} and
\ref{nz_B}.  Ideally, one should use the incompleteness calculated for
each spectral class. However, galaxies with no redshift measurement
have \emph{no} spectral-type determination. We have examined the
dependence of the incompleteness as a function of galaxy-type using
the colors of the galaxies without redshift, as these are correlated
with galaxy type. For the $R_\mathrm{c}\le20.5$ sample, the
incompleteness is uniform with galaxy colors, justifying the use of
the average incompleteness for the full sample.  For the $V\le21.0$
and $B\le22.0$ samples, the incompleteness is significantly stronger
for the bluer galaxies; we cannot however evaluate the incompleteness
per spectral-type from the colors as the relation between color and
spectral-type suffers a large dispersion. The relative larger
incompleteness in blue galaxies at faint magnitudes of the $V$ and
$B$ spectroscopic samples converts into a relative larger
incompleteness in faint late-type galaxies compared to early-type
galaxies. This might explain why the expected curves for the late-type
galaxies in \fgs~\ref{nz_V} and \ref{nz_B} appear to systematically
under-estimate the observed distribution.

As already mentioned in \sct~\ref{sp_kcor}, the ESS late-type class
contains predominantly Sc+Sd and dI galaxies.  An interesting issue is
how the giant (Sc+Sd) and dwarf (dI) galaxies contribute to this
evolution.  In the composite fit to the late-type LF at
$R_\mathrm{c}\le21.5$ shown in the lower left panel of
\fg~\ref{lf_local_RVB}, the giant and dwarf components cross-over near
$M(R_\mathrm{c})\simeq-18.5+5\log h$ for
$(\Omega_m,\Omega_\Lambda$)=(1.0,0.0); for
$(\Omega_m,\Omega_\Lambda$)=(0.3,0.7), this magnitude converts
approximately to $M(R_\mathrm{c})\simeq-18.8+5\log h$ (see
\sct~\ref{shape}).  We therefore calculate the observed and expected
redshift distributions for the 2 following sub-samples: the bright
sample, with $M(R_\mathrm{c})\le-19.0+5\log h$, containing 227
galaxies, which are predominantly Sc+Sd galaxies; the faint sample,
with $M(R_\mathrm{c})\ge-19.0+5\log h$, containing 82 galaxies, which
are predominantly dI galaxies. Lower panel of \fg~\ref{nz_ScIm} shows
that an evolution factor with $P_{0.15}=3.0$ has a small impact on the
distribution of galaxies with $M(R_\mathrm{c})\ge-19.0+5\log h$, and
both expected curves are compatible with the observed distribution. In
contrast, the expected no evolution curve provides a poor match to the
observed distribution of galaxies with $M(R_\mathrm{c})\le-19.0+5\log
h$, whereas the expected curve with $P_{0.15}=3.0$ provides a better
agreement.  The expected distribution for a higher evolution rate,
$P_{0.15}=6.0$, is also shown in \fg~\ref{nz_ScIm} for the bright
galaxies: it provides an even better adjustment (note that if the dI
galaxies where not evolving, a higher evolution rate than measured for
the Sc+Sd+dI altogether would be expected for the Sc+Sd galaxies
alone).  This suggests that the Sc+Sd galaxies are likely to
contribute significantly to the detected late-type evolution. We
however cannot exclude evolution of the dI galaxies.

\section{Comparison of the ESS evolution with other surveys \label{evol_comp}}

Most other redshift surveys to $z\sim0.5-1.0$ detect evolution in the
luminosity function. The evolution affects either the amplitude $\Phi$
of the LF (number-density evolution), or its shape via the
characteristic magnitude (luminosity evolution) and/or its faint-end
behavior. Evolution in the luminosity density is also often used for
measuring the evolution rate, as it has the advantage to account for
the 3 types of evolution.  Note that for a non-evolving shape of the
luminosity function $\varphi(M)\mathrm{d}M$ , any evolution in the
amplitude $\Phi$ (see \eq~\ref{schechter2}) yields an identical
evolution rate of the luminosity density.

In the following, we only consider the evidence for separate evolution
of the E/S0 and Spiral galaxies, as evolution in the full galaxy
population does not allow one to isolate the evolving population.  For
instance, the total luminosity density $\rho_L$ in the CFRS shows an
increase with redshift which \citet{lilly96} model as
$\rho_L(z)\propto(1+z)^{2.7\pm0.5}$ at 4400 \AA~for an
($\Omega_m$,$\Omega_\Lambda$)=(1.0,0.0) cosmology. This is close to
the value $\gamma=2.6$ obtained from the ESS $B$ band number-counts in
Table~\ref{evol_BVR}, suggesting that the evolution rate for the
Spiral galaxies in the CFRS may be higher than in the ESS. A direct
measure of the evolution rate of the blue galaxies in the CFRS (bluer
than a non-evolving Sbc galaxy) would however be required for a
quantitative comparison with the ESS.

Because there is weak evidence for evolution in the faint-end slope of
the LF in the existing surveys \citep{heyl97}, we restrict the
following discussion to the evidence for (i) number-density evolution,
and (ii) luminosity evolution.

\subsection{Number-density evolution     \label{dens_comp} }  

\citet{ellis96} detect a marked density evolution in the Autofib
star-forming galaxies by a factor 2 between $z\sim0.15$ and
$z\sim0.4$, which would correspond to an overall fading of the
population by $0.5$\mag~ in the $b_J$ band. This is comparable to the
ESS evolution rate derived above: for $P_{0.15}=3.0$, the density
increases by a factor $1.75$ between $z\sim0.15$ and
$z\sim0.4$. Further analysis of the Autofib survey based on galaxy
spectral types \citep{heyl97} leads to detection of a strong density
evolution in the late-type spiral galaxies (Sbc and Scd) which they
model as $\phi^*(z)\propto(1+z)^{2.9}$ for the Sbc galaxies and as
$\phi^*(z)\propto(1+z)^{3.5}$ for the Scd galaxies in a cosmology with
($\Omega_m$,$\Omega_\Lambda$)=(1.0,0.0). These values are in
acceptable agreement with the $\gamma=2.6$ value derived from the ESS
in the $B$ band.

In the CNOC2, the dominant evolution detected by \citet{lin99} in the
redshift range $0.1\la z\la 0.55$ is a strong density evolution of the
late-type galaxies, defined as galaxies with
$UBVR_\mathrm{c}I_\mathrm{c}$ colors similar to those computed from
the Scd and Im templates of \citet{coleman80}; the authors model the
evolution as $\phi^*(z)\propto e^{Pz}$, with $P=2.8\pm1.0$ and
$P=2.3\pm1.0$ in the $B_{AB}$ band for $\Omega_m=1.0$, $\Omega_m=0.2$
\resp; and $P=2.9\pm1.0$, $P=2.6\pm1.0$ in the $R_\mathrm{c}$ band for
$\Omega_m=1.0$, $\Omega_m=0.2$ respectively.  At $z\la0.1$, $e^{Pz}$
and $(1+z)^\gamma$ can be expanded into $1+Pz$ and $1+\gamma z$ \resp,
which would yield a reasonable agreement at low redshift between the
CNOC2 and the ESS in both the $B$ and $R_\mathrm{c}$ bands.  However,
at $z\sim1$, $e^{2.5z}$ is a factor of 2 larger than $(1+z)^{2.5}$,
implying a stronger evolution rate in the CNOC2 than in the ESS.

Obtained with a similar observational technique as the CNOC2, the
field sample of the CNOC1 \citep{lin97} shows an increase of the
luminosity density of galaxies by a factor 3 between $z\sim0.25$ and
$z\sim0.55$ for galaxies with rest-frame colors bluer than a
non-evolving Sbc galaxy. In the NORRIS survey of the Corona Borealis
Supercluster, \citet{small97b} detect a similar evolution rate: the
amplitude $\phi^*$ of the Schechter LF for galaxies with strong [OII]
emission line increases by nearly a factor 3 from $z\la0.2$ to $0.2\le
z\le0.5$. These various values of the number-density evolution rate
are stronger than the increase by a factor $1.8$ in the ESS density of
late-type galaxies which we derive for $P_{0.15}=3.0$ between
$z\sim0.2$ et $z\sim0.5$.

In the CADIS survey, in which redshifts are derived from a combination
of wide and medium-band filters, \citet{fried01} detect an increase of
the Johnson $B$ luminosity density of Sa-Sc galaxies with redshift,
which is partly due to an increase in the amplitude $\phi^*$ of the
fitted Schechter luminosity function, and can be modeled as
$\rho(z)\propto(1+6.9z)$ for ($\Omega_m$,$\Omega_\Lambda$)=(1.0,0.0),
and as $\rho(z)\propto(1+0.78z)$ for
($\Omega_m$,$\Omega_\Lambda$)=(0.3,0.7). The evolving term in the
luminosity density can be converted into
$\rho(z)\propto[1+3.4(z-0.15)]$,
\resp~$\rho(z)\propto[1+0.70(z-0.15)]$. Whereas the CADIS evolution
rate is similar to that in the ESS for
($\Omega_m$,$\Omega_\Lambda$)=(1.0,0.0), it is much smaller than in
the ESS for ($\Omega_m$,$\Omega_\Lambda$)=(0.3,0.7) (see
Table~\ref{evol_BVR}); note however that this comparison may be
complicated by the fact that the considered evolving CADIS population
contains early-type Spiral galaxies, to the contrary of the ESS
late-type spectral class.

In contrast, the recent COMBO-17 survey \citep{wolf03} which is also
based on a combination of wide and medium-band filters, shows no
significant evolution in the number-density and luminosity density of
either Sa-Sc and Sbc-Starburst galaxies from $z\sim0.3$ to $z\sim1.1$
in the Johnson $B$ and SDSS $r$ bands \citep{fukugita96}.  Moreover,
whereas the various mentioned surveys (CFRS, Autofib, NORRIS, CNOC2,
CNOC1, CADIS) show no or a weak change in the luminosity density of
early-type (E-S0) or red galaxies over the considered redshift range,
the COMBO-17 survey detects a marked increase with redshift by a
factor of 4 in the contribution from the E-Sa galaxies to the $r$ and
$B$ luminosity densities for ($\Omega_m$,$\Omega_\Lambda$)=(0.3,0.7)
\citep{wolf03}. The different results between the COMBO-17 and the
other redshift surveys may be due to the complex selection effects
inherent to surveys based on multi-medium-band photometry such as the
COMBO-17, and which are most critical for emission-line
galaxies. These effects however do not seem to affect the CADIS
survey.

At last, the evolution detected in the far infrared from IRAS galaxies
\citep{saunders90,bertin97b,takeuchi03} can be characterized as
$\phi^*(z)\propto(1+z)^\gamma$ with $\gamma\simeq3-3.4$ for pure
density evolution (both cosmologies considered in this article are
used, depending on the authors). The evolution of the IRAS galaxies is
consistent with the ESS late-type evolution, in agreement with the
fact that IRAS galaxies may represent a sub-population of the optical
spiral galaxies.

\subsection{Luminosity evolution              \label{lum_comp} }  

The apparent density evolution detected in the ESS could also be
produced by a luminosity evolution of the late-type spiral galaxies:
if these galaxies were brighter at higher redshift, they would enter
the survey in larger numbers at a given apparent magnitude.  Using the
values of the power-law index $\gamma$ listed in Table~\ref{evol_BVR}
and the slopes of the ESS magnitude number-counts \citep{arnouts97},
we can estimate a corresponding magnitude brightening.  For
cosmological parameters ($\Omega_m$,$\Omega_\Lambda$)=(1.0,0.0), we
measure from Table~\ref{evol_BVR} an increase in the number-density of
galaxies by a factor $2.8$ between $z\simeq0$ and $z\simeq0.5$, and by
a factor $5.7$ between $z\simeq0.5$ and $z\simeq1$ in the
$R_\mathrm{c}$ band; in the $B$ bands, the density increases by a
factor $2.9$ at $z\simeq0.5$, and $6.1$ at $z\simeq1$.  For
($\Omega_m$,$\Omega_\Lambda$)=(0.3,0.7), the density increases by
$2.1$ and $3.5$ at $z\simeq0.5$ and $z\simeq1$ \resp~in
$R_\mathrm{c}$, and by $2.4$ and $4.6$ at $z\simeq0.5$ and $z\simeq1$
\resp~in $B$.  Using the slopes $\beta=0.38$ in the $R_\mathrm{c}$
band and $\beta=0.46$ in the $B$ band ($n(m)\propto10^{\beta m}$) for
the ESS magnitude number-counts \citep{arnouts97}, these values of the
density increase are equivalent to $\sim1.2$\mag~and
$\sim2.0$\mag~brightening of the late-type ESS galaxies at
$z\simeq0.5$ and $z\simeq1$ \resp~in $R_\mathrm{c}$, and a
$\sim1.1$\mag~and a $\sim1.7$\mag~brightening \resp~in $B$ for
($\Omega_m$,$\Omega_\Lambda$)=(1.0,0.0); and to $\sim0.8$\mag~ and
$\sim1.4$\mag~brightening \resp~in both the $R_\mathrm{c}$ and $B$
bands for ($\Omega_m$,$\Omega_\Lambda$)=(0.3,0.7) (the steeper
evolution rate in the $B$ band is compensated by a steeper slope of
the number-counts also in the $B$ band).



These brightening estimates for the ESS late-type galaxies with
($\Omega_m$,$\Omega_\Lambda$)=(0.3,0.7) are comparable to those caused
by the passive evolution of an Sc galaxy (due to the evolution of the
stellar population). From the model predictions of \citet[][using
($\Omega_m$,$\Omega_\Lambda$)=(0.45,0.0) and $H_0 = 50$ km s$^{-1}$
Mpc$^{-1}$, which imply an age of the Universe of 15
Gyr]{poggianti97}, an Sc galaxy brightens by $\sim0.6$\mag~and
$\sim1.2$\mag~in its rest-frame $R$ band at $z\sim0.5$ and $z\sim1.0$
\resp, and by $\sim0.9$\mag, $\sim1.5$\mag~\resp~in its rest-frame $B$
band. A comparable or stronger brightening is expected for the
Elliptical galaxies at these redshifts: $\sim0.6$\mag~ and
$\sim1.4$\mag~ in $R$, $\sim0.7$\mag~ and $\sim3.0$\mag~ in $B$
\citep{poggianti97}.  However, there are indications of a marked
decrease in the number density of E galaxies with redshift
\citep{fried01,wolf03}, which compensates for their luminosity
evolution, which explains why no evolution in this population is
detected by the ESS.

\citet{lilly95} detect a 1\mag~brightening of the CFRS blue galaxies
(defined as galaxies with rest-frame colors bluer than a non-evolving
Sbc template from \citealt{coleman80}) between the intervals $0.2\la
z\la0.5$ and $0.5\la z\la0.75$ in an
($\Omega_m$,$\Omega_\Lambda$)=(1.0,0.0) cosmology; they however cannot
discriminate whether this brightening is due to luminosity or density
evolution. With $\gamma=2.6$ (see Table~\ref{evol_BVR}), the ESS $B$
density of late-type galaxies increases by a factor $1.6$ between
$z\simeq0.25$ and $z\simeq0.625$ (the median value of the 2 quoted
CFRS intervals), which corresponds to a brightening of $\sim0.44$\mag,
nearly a factor 2 smaller than in the CFRS.  \citet{cohen02} also
detect for the emission-line dominated galaxies at $z\sim1$, when
compared to the measurement of \citet{lin96} at $z\sim0$, a mild
brightening by $\sim0.75$\mag~in the $R$ band for an
($\Omega_m$,$\Omega_\Lambda$)=(0.3,0.0) cosmology, which is a factor 2
smaller than in the ESS estimated brightening in the $R_\mathrm{c}$
band for ($\Omega_m$,$\Omega_\Lambda$)=(0.3,0.7).

\section{Conclusions, further discussion and prospects         \label{concl} } 

Using the Gaussian+Schechter composite LFs measured for the
ESO-Sculptor Survey, we obtain evidence for evolution in the late
spectral-type population containing late-type Spiral (Sc+Sd) and dwarf
Irregular galaxies.  This evolution is detected as an increase of the
galaxy density $n(z)$ which can be modeled as
$n(z)\propto1+P_{0.15}(z-0.15)$ with $P_{0.15}\sim3\pm1$ or as
$n(z)\propto(1+z)^\gamma$ with $\gamma\sim2\pm1$ using the currently
favored cosmological parameters
($\Omega_m$,$\Omega_\Lambda$)=(0.3,0.7); for
($\Omega_m$,$\Omega_\Lambda$)=(1.0,0.0), $P_{0.15}\sim4\pm1$ and
$\gamma\sim2.5\pm1$.  Both models yield a good match of the ESS
$BVR_\mathrm{c}$ redshift distributions to $21-22$\mag~and the
number-counts to $23-23.5$\mag, which probe the galaxy distribution to
redshifts $z\sim0.5$ and $z\sim1.0$ respectively.  Using \emph{both}
the redshift distributions and the number-counts allows us to lift
part of the degeneracies affecting faint galaxy number counts: the
redshift distributions allow to isolate the evolving populations,
whereas the faint number counts provide better constraints on the
evolution rate. These results are based on the hypothesis that the
shape of the LF for the ESS late-type class does not evolve with
redshift out to $z\sim1$.

Examination of the other existing redshift surveys to $z\sim0.5-1.0$
indicates that a wide range of number-density evolution rates have
been obtained. The evolution rate of the Sc+Sd+dI galaxies detected in
the ESS is among the range of measured values, with some surveys
having weaker of higher evolution rates.  The most similar survey to
the ESS, the CNOC2, yields a twice larger increase in the number
density of late-type Spiral and Irregular galaxies at $z\sim1$. 

A priori, density evolution indicates that mergers could play a
significant role in the evolution of late-type Spiral and Irregular
galaxies.  \citet{lefevre00} detect a $\sim20$\% increase in the
fraction of galaxy mergers from $z\sim0$ to $z\sim1$, which can be
modeled as $\propto(1+z)^{3.2}$; interestingly, examination of their
\fg 1 indicates that a significant fraction of the merger galaxies
have a Spiral or Irregular structure.

The ESS density increase for the Sc+Sd+dI galaxies could also be
caused by a $\sim1$\mag~brightening of these galaxy populations at
$z\sim0.5$ and a $\sim1.5-2.0$\mag~brightening at $z\sim1$ (depending
on the filter and cosmological parameters). This luminosity evolution
is compatible with the expected passive brightening of Sc galaxies at
increasing redshifts \citep{poggianti97}.  \citet{driver01} also shows
that the Hubble Deep Field \citep{williams96} bi-variate brightness
distributions for Elliptical, Spiral, and Irregular galaxies are all
consistent with passive luminosity evolution in the 3 redshift bins
$0.3-0.6$, $0.6-0.8$, $0.8-1.0$.  The ESS brightening at $z\sim1$
agrees with the value measured from the CFRS blue galaxies
\citep{lilly95}, but is twice smaller than that measured by
\citet{cohen02} for emission-line dominated galaxies.

In all analyses of the redshift and magnitude distributions, the major
difficulty is to distinguish between luminosity and density evolution,
as these produce the same net effect on the redshift and magnitude
distributions.  Interpretation of density and luminosity evolution of
a galaxy population is also complicated by possible variations in the
star formation rate with cosmic time: \citet{lilly98} evaluate an
increase in the star formation rate of galaxies with large disks by a
factor of $\sim3$ at $z\sim0.7$, which shows as an increase of the
luminosity density at bluer wavelengths. Using PEGASE \citep{fioc97},
\citet{rocca99} also show that the Sa-Sbc galaxies have a star
formation rate which varies more rapidly in the interval $0\la z\la 1$
than for the E/S0 or Sc-Im galaxies.

The ESS suggests that the Sc+Sd galaxies are an evolving population,
but evolution in the dI galaxies cannot be excluded. Whether the
Spiral galaxies, or the Irregular galaxies, or both populations
contribute significantly to the excess number-counts is still unclear
from the various existing analyses: using photometric redshifts,
\citet{liu98} detected a significant amplitude increase and
brightening of the $U$ LF of Sbc and bluer galaxies in the redshift
interval $0\la z \la0.5$, together with an excess population of
starburst galaxies at $z\ga0.3$ which are absent at $z\la0.3$; galaxy
number counts in the near infrared \citep{martini01,totani01}, which
are less sensitive to current star formation, also allow some moderate
number evolution of the Spiral galaxies and/or Irregular galaxies
(with $\gamma\sim1$, see above).  Evolutionary effects are also
detected at higher redshifts: \citet{driver98} show that at $1\la z
\la3$, the redshift distributions of Sabc galaxies to $I>25$ and of
Sd/Irr galaxies to $I>24$ require some number and luminosity evolution
\citep[but see][]{driver01}. This could be consistent with the recent
deep optical and infrared observations which favor mild luminosity
evolution of the overall galaxy population
\citep{pozzetti03,kashikawa03}. Nevertheless, none of these surveys
allow one to isolate one evolving population among the Spiral and
Irregular/Peculiar galaxies.

In contrast, some surveys favor Irregular/Peculiar galaxies as a major
contributor to the excess count: an excess population of
Irregular/Peculiar galaxies was directly identified as the cause for a
strong deviation from no-evolution in the $I$ and $K$ number-counts per
morphological type \citep{glazebrook95,huang98}; a population of
gently-evolving starbursting dwarves was also invoked to explain these
excess objects \citep{campos97a}. Using morphology of galaxies
obtained from Hubble Space Telescope images, \citet{im99} provide
further evidence for a marked increase in the relative abundance of
Irregular and Peculiar galaxies which they interpret as starbursting
sub-$L^*$ E/S0 and Spiral galaxies. \citet{totani98} also detect such
an excess population at $z\ga0.5$. Using again Hubble Space Telescope
imaging for galaxy morphology, \citet{brinchmann98} estimate a $\sim
30$\% increase in the proportion of galaxies with an irregular
morphology at $0.7\la z\la0.9$ \citep[see
also][]{abraham96,volonteri00}.

Other surveys identify Spiral galaxies as the evolving population:
morphological number counts based on Hubble Space Telescope images
indicate that the Spiral counts rise more steeply than the
no-evolution model \citep{abraham96}. By complementing space-based
images with ground-based spectroscopic redshifts, \citet{brinchmann98}
detect a $\sim1$\mag~brightening of the Spiral galaxies by $z\sim1$.
\citet{schade96} also measure a $\sim1.6$\mag~brightening of the
central surface brightness of galaxy disks at $0.5\la z \la1.1$.

We emphasize that measuring a reliable evolution rate requires a
realistic parameterization of the intrinsic LFs of each galaxy
population. We show here that the ESS Gaussian+Schechter composite LFs
provide more robust constraints on the evolution rate than pure
Schechter LFs, as a small change in the faint-end slope has a large
incidence onto the number-counts. In contrast, variations in the
faint-end slope of the dE and dI galaxies, the only populations for
which the Schechter faint-end is poorly determined have a smaller
impact on the adjustment on the number counts.  \citet{totani01} also
show that separating the E and dE galaxies into a Gaussian+Schechter
LF does yield a better agreement of the $K$ number-counts at
$K\ga22.5$ than a pure Schechter function for the joint class of E+dE.

By using better measures of the intrinsic LFs for the various galaxy
types, one should be able to obtain improved measurements of the
galaxy evolution rates, and eventually to use the faint number counts
to probe the cosmological parameters. \citet{koo93} and
\citet{gronwall95} had already suggested that passive luminosity
evolution is sufficient to match the $B$ number counts, thus implying
that the number counts could be used to probe the curvature of space.
\citet{pozzetti96} and \citet{metcalfe01} then showed that passive
luminosity evolution with a low value of $\Omega_m$ provide good fits
to UV-optical-near-infrared galaxy counts.  The recent adjustments of
very deep optical and near-infrared galaxy counts
\citep{totani00,nagashima01,totani01,nagashima02} which probe the
galaxy distribution to $z\sim3$ further confirm that the Einstein-de
Sitter cosmology with ($\Omega_m$,$\Omega_\Lambda$)=(1.0,0.0) is
excluded at a high confidence level \citep[see also][]{totani97,he00};
whereas by using the currently favored values of $\Omega_m=0.3$ and
$\Omega_\Lambda=0.7$, these fits constrain the evolution rate of
galaxies in the hierarchical clustering picture, with only some mild
number evolution of Sbc/Sdm galaxies allowed \citep{totani01}.

Under-going deep redshift surveys to $z\ga1$ raise new prospects along
this line. The present ESS analysis shows the usefulness of using both
the magnitude \emph{and} redshift distributions for studying galaxy
evolution.  By obtaining the redshift distributions per galaxy type to
$z\ga1$ over large volumes which average out the large-scale
structure, spectroscopic redshift survey such as the VIRMOS
\citep{lefevre03b} and DEEP2 \citep{davis03} projects should provide
improved clues on the evolving galaxy populations at $z\sim1$ and
better constrain the nature of this evolution.  If these surveys
confirm that all galaxy types only experience mild luminosity/density
evolution, complementing the deep infrared galaxy counts with
morphological information might allow one to lift the degeneracy in
the faint infrared number-counts and to confirm whether a low
matter-density universe is favored. This will however require a
detailed knowledge (i) of the luminosity-size relation for distant
galaxies of various morphological type (used to model the selection
effects caused by the cosmological dimming in surface brightness, see
\citealt{totani00}), and (ii) of the interstellar and intergalactic
extinction.
 
\begin{acknowledgements}

We are grateful to Eric Slezak for providing his programmes for
calculation of the selection functions and cosmological distances.
G. Galaz acknowledges the support of FONDAP grant \#15010003 "Center
for Astrophysics". We also thank the anonymous referee whose comments
helped in improving the presentation of this article.

\end{acknowledgements}

\bibliography{0140}
\bibliographystyle{aa}

\end{document}